\documentclass[preprint,preprintnumbers,superscriptaddress,amsmath]{revtex4}

\usepackage[dvips]{graphics}
\usepackage{bm}
\usepackage{psfrag}
\usepackage[psamsfonts]{amssymb}
\usepackage[T1]{fontenc}
\usepackage{graphicx}
\usepackage{tabularx}

\newcommand{\be}{\begin{equation}}
\newcommand{\ee}{\end{equation}}
\newcommand{\ba}{\begin{eqnarray}}
\newcommand{\ea}{\end{eqnarray}}

\begin{document}

\preprint{APS preprint}

\title{The Dynamics of Book Sales: Endogenous versus Exogenous Shocks in
Complex Networks}

\author{F. Desch\^atres}
\affiliation{Ecole Normale Sup\'erieure, rue d'Ulm, Paris, France}

\author{D. Sornette}
\affiliation{Institute of Geophysics and Planetary Physics
and Department of Earth and Space Sciences,
University of California, Los Angeles, CA 90095}
\affiliation{Laboratoire de Physique de la Mati\`ere Condens\'ee,
CNRS UMR 6622 and Universit\'e de Nice-Sophia Antipolis, 06108
Nice Cedex 2, France}
\email{sornette@moho.ess.ucla.edu}

\date{\today}

\begin{abstract}

We present an extensive study of the foreshock and aftershock signatures
accompanying peaks of book sales. The time series of book sales
are derived from the ranking system of
Amazon.com. We present two independent ways of classifying peaks, one based
on the acceleration pattern of sales and the other based on the exponent
of the relaxation. They are found to be consistent and reveal the
co-existence of two types of sales peaks:
exogenous peaks occur abruptly and are followed by
a power law relaxation, while endogenous sale peaks occur after a
progressively accelerating power law growth followed by an approximately
symmetrical power law relaxation which is slower than for exogenous
peaks.  We develop
a simple epidemic model of buyers connected within a network of
acquaintances which propagates rumors and opinions on books.
The comparison between the predictions of the model and
the empirical data confirms the validity of the model and
suggests in addition that social networks have evolved to
converge very close to criticality (here in the sense of critical
branching processes of opinion spreading). We test in details the evidence
for a power law distribution of book sales and confirm a previous
indirect study suggesting that the fraction of books (density distribution)
$P(S)$ of sales $S$
is a power law $P(S) \sim 1/S^{1+\mu}$ with $\mu \approx 2$.

\end{abstract}


\maketitle

\section{Introduction}

In the context of linear response theory, the fluctuation
dissipation theorem provides an explicit relationship between
microscopic dynamics at equilibrium and the macroscopic response
that is observed in a dynamic measurement. It relates equilibrium
fluctuations to close-to-equilibrium observables. In
out-of-equilibrium systems, the existence of a relationship between
the response function to external kicks and spontaneous internal
fluctuations is not settled \cite{ruelle}. In many complex systems,
this question amounts to distinguishing between endogeneity and
exogeneity and is important for understanding the relative effects
of self-organization versus external impacts. This is difficult in
most physical systems because externally imposed perturbations may
lie outside the complex attractor which itself may exhibit
bifurcations. Therefore, observable perturbations are often
misclassified. It is thus interesting to study other systems, in
which the dividing line between endogenous and exogenous shocks may
be clearer in the hope that it would lead to insight about complex
physical systems. The systems for which the endogenous-exogenous
question is relevant span, beyond the physical sciences, the
biological to the social sciences \cite{reviewsorendoexo}.

Here, we study a real-world example of how the response function to external
kicks is related to internal fluctuations.
We study the precursory and recovery signatures accompanying
shocks in complex networks, that we test on the database of book ranks
provided by Amazon.com. We find clear distinguishing
signatures classifying two types of sales peaks. Exogenous peaks
occur abruptly and are followed by a power law relaxation, while
endogenous sales peaks occur after a progressively accelerating
power law growth followed by an approximately symmetrical power
law relaxation which is slower than for exogenous peaks. These
results are rationalized quantitatively by a simple model of
epidemic propagation of interactions with long memory within a
network of acquaintances. The slow relaxation of sales implies
that the sales dynamics is dominated by cascades rather than by
the direct effects of news and advertisements, indicating that the
social network is close to critical. We perform also a direct
measurement on ranks that give important constraints on the
conversion that transforms sales into ranks. A short version
of this work is \cite{endoPRL}.

The structure of the paper is as follows.
We first present our database (Junglescan.com) and explain how we use it
to estimate the sales from the ranks. To do so, we need to make some
assumptions about the mechanisms underlying the sales dynamics. We
discuss these points on section \ref{Converting ranks into sales}. In
section \ref{Direct measure on rank}, we revisit
this question and derive a way to measure
directly this conversion.
The epidemic branching process used in our study leads to an effective
linear coarse-grained description of the complex nonlinear dynamics. We
present our model on section \ref{Description of the Model} and derive
some predictions about the behavior of the system before and after sales peaks.
Then, the data are analyzed in section \ref{sdth}. We provide
two independent classifications of exogenous and
endogenous shocks. The obtained
classifications are robust. The last section concludes with suggestions for
future research.

\section{Construction of the data: converting ranks into sales}
\label{Converting ranks into sales}

  \subsection{The database}

The study was performed using data from Amazon.com. Amazon was
founded in 1994 as an online bookseller, and since has expanded
its business into areas such as clothing, gourmet food, sports
equipment and jewelry. With an expected $\$6$ billion net sales in
2004, the American giant of internet trading is by far the largest
e-retailer.

Electronic data make it possible to deal with huge amount of information
impossible to gather otherwise. With the birth of the Internet, the
prospect of understanding quantitatively social
phenomena is emerging \cite{Roehneretal}.
It is perhaps, the first time that an organization can collect so much
information\footnote{Of course, we should mention search engines such as
Google. To have a look at most popular queries, look at
http://www.google.com/press/zeitgeist.html.}. What people buy can help
give an image of the state of the society. And Amazon can provide almost
all of what can be bought. It operates on a global scale. That is why
Amazon is an outstanding means to probe the society. As Andreas S.
Weigend, chief scientist of Amazon.com (2002-2004) wrote on his webpage:
``Amazon.com might be the world's largest laboratory to study human
behavior and decision making.''
Unfortunately, for obvious competition grounds, Amazon has no
incentive to share its data. It jealously keeps its sales secret.
But, Amazon is not completely unknowable. It provides on its
website the rank for each of its product based on their passed sales.

In a first step, Y. Ageon in our group developed an application using
``WebBrowser control'' to capture automatically the ranks of books on
Amazon.com at regular time intervals. WebBrowser
control\footnote{http://msdn.microsoft.com/library/default.asp?url=/workshop/browser/webbrowser/webbrowser.asp}
allows a user to browse sites on the Internet's
World Wide Web. The application first opens
Amazon's page with the URL : http://207.171.185.16/exec/obidos/ASIN/XXXXXX
where XXXXXX corresponds to the ISBN or ASIN code
of the targeted product (book,
DVD, CD, Game, etc.). The rank is then found on
the page using an algorithm which
processes character string. In this way, we constructed a database of the ranks
of tens of books over several months with a sampling periodicity of one hour.
This allowed us to check the quality of the time
series of the ranks of thousands of
books which have been recorded by \textbf{JungleScan}
(http://www.junglescan.com) over several years.
JungleScan scans books on Amazon
typically every six hours for those updated hourly by Amazon. For all books
we have checked, we found identical rank values
in \textbf{JungleScan} and in our
directly constructed database, showing that we could trust the data
from \textbf{JungleScan} for our study.

   \subsection{Amazon ranking schemes}
   \label{ranking schemes}

In order for this data to be useful, we need to
convert the ranks into meaningful
``physical'' units, such as sales or sale rates. The problem is that
Amazon.com does not divulge the exact formula for this conversion (otherwise,
its secretive strategy on its sale figures would be useless). In our study,
we use the time series of book ranks up to April 2004. Until October 2004,
the following ranking system held, which is the relevant one
for our study\footnote{In October 2004, Amazon completely redesigned its
ranking system. See
http://www.fonerbooks.com/surfing.htm for details.
Not only the meaning of the ranks have changed, due to the inclusion
of e-books and Marketplace sales, but also the infrastructure of the
system has been modified: the three tiers are gone, all ranks are updated every cycle, which
runs around once an hour. It is thus no more possible to compare
the ranks before to those after October 2004. It is important to 
realize that Amazon's ranking system may again evolve on a moment's notice.}.

        \subsubsection{The official statement}
            \label{official statement}
             Amazon gives some hints about its ranking system.
             The official statement of Amazon is the following :

                     \begin{quotation}

                     \textbf{What Sales Rank Means}

                     As an added service for
customers, authors, publishers, artists,
                     labels, and studios, we show
how items in our catalog are selling.
                     This bestseller list is much
like The New York Times Bestsellers
                     List, except it lists
millions of items! The lower the number, the
                     higher the sales for that particular title.

                     The calculation is based on Amazon.com sales and is updated
                     regularly. The top 10,000 best sellers are updated each
                     hour to reflect sales in the
preceding 24 hours\footnote{We indeed
                     found a strong spectral peak
at period $1$ day on the spectrum of the
                     time series of book ranks.}. The next 100,000 ranks
                     are updated daily. The rest
of the list is updated monthly, based
                     on \textsl{several different factors}.
                     \end{quotation}

          \subsubsection{Different time scales}
                     Amazon can use different time scales :

                     \begin{itemize}
                     \item a short time scale: one
day. That is the time scale we are
                     interested in since our purpose is to study sales dynamics.

                     \item a long time scale. For
instance, if Harry Potter sales were
                     to crash overnight, its rank
won't fall so sharply. It means that
                     the whole history of a book
can outweigh its instantaneous sales.
                     It is worth understanding when the short time scale is
                     outbalanced.
                     \end{itemize}

             \subsubsection{Evidence that Amazon uses different ranking schemes}

            Books with ranks in the range below
$100,000$ are re-ranked according to
            the sales during the last $24$ hours. For these books, Amazon
            uses a short time scale to rank them. We will see that there
            are  exceptions. For books with sales
ranks over $100,000$, Amazon does not
            explain the different factors used to update the ranks.

\begin{figure}
                \begin{center}
                  \includegraphics[width=.6\textwidth]{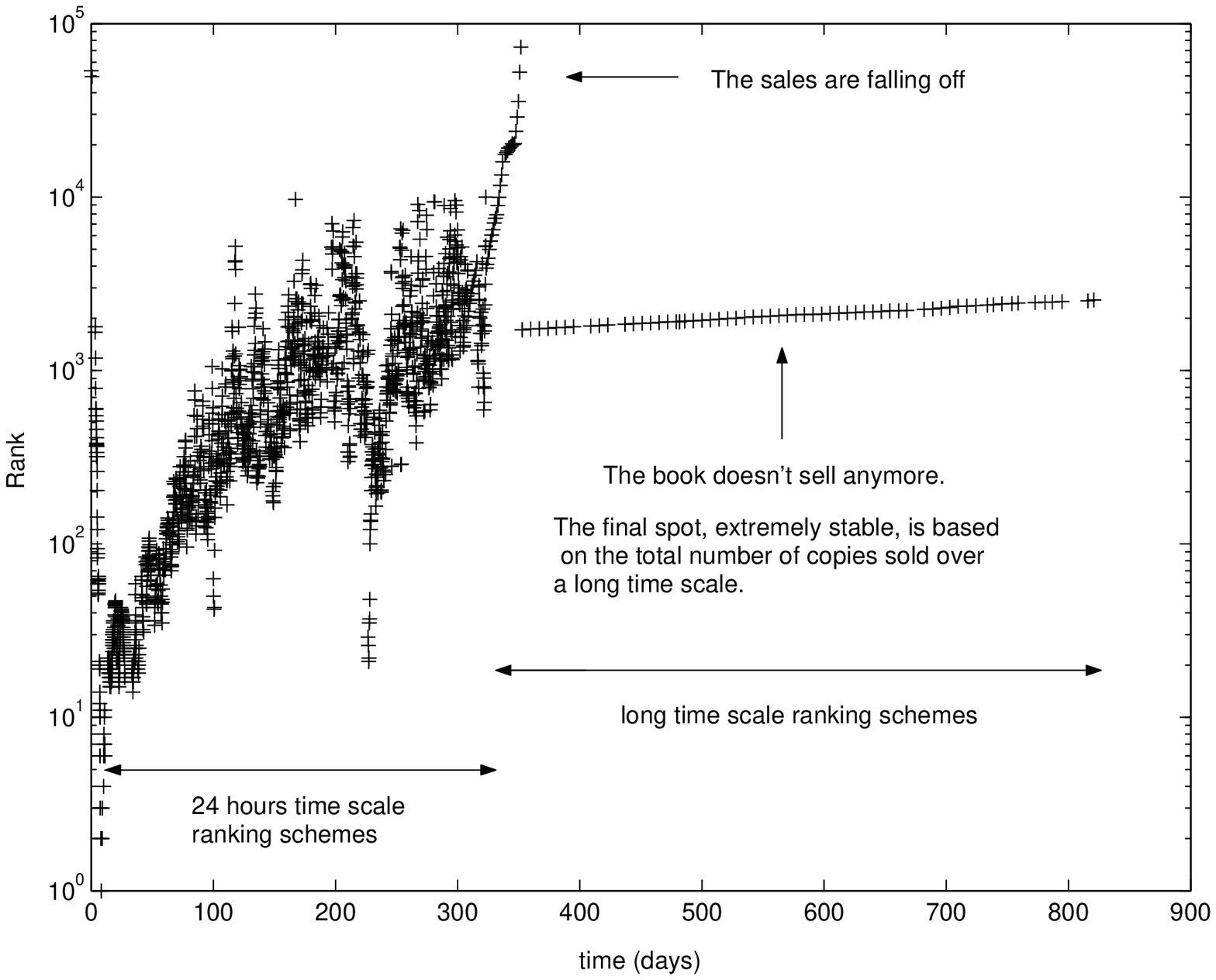}
                \end{center}
                \begin{center}
                  \includegraphics[width=.65\textwidth]{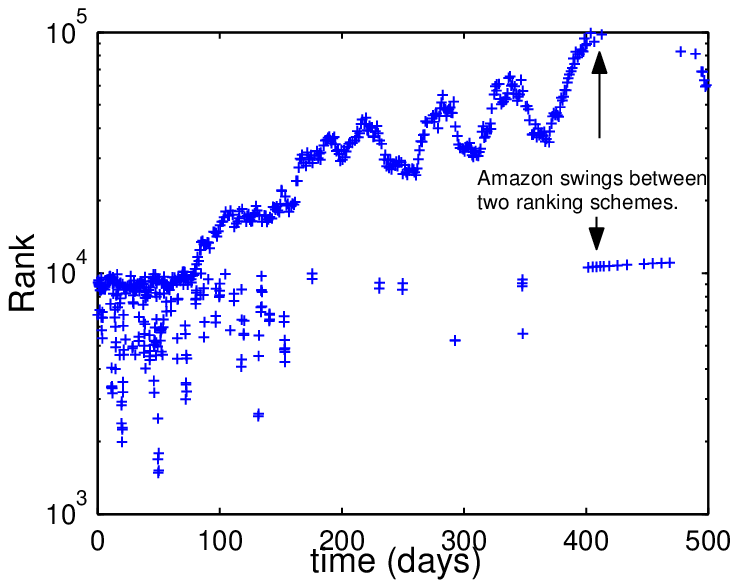}
                  \caption{Two time series of
ranks showing that Amazon switches between at
                  least two ranking schemes, as explained in the text.
                  The first book (top) is ``See No
Evil'' by R. Baer. The second book
                  (bottom) is ``F'd Companies'' by P. Kaplan.}
             \label{rank}
             \end{center}
  \end{figure}

            Fig.\ref{rank} (top panel) shows an example in which the rank
            increased with large fluctuations up to $100,000$ (the sales were
            steadily decreasing), then
            jumped to a few thousands and then
followed a very smooth slowly increasing trajectory.
            The absence of fluctuations in the
second rightmost part of the graph
            implies that Amazon re-ranked the books using its sales
            data using a long time-scale.
            Fig.\ref{rank} (bottom panel) shows a related effect.
            First, the sales fell off resulting in the rank
            increasing to $100,000$. At this rank
level, Amazon switched to a long time scale
            ranking scheme, leading to a
reassessment of the rank in the range of $10,000$.
            At some later time, some sales occurred which led Amazon to
            switch back to a $24$ hours time scale ranking scheme, resulting in
            the rank increasing dramatically to a
level around $10^5$ but slightly below
            (so that the $24$ hours ranking system is active).

           To sum up, only books with a low rank (typically less
            than ten thousands) are ranked at a
time scale from a few hours to a day and
            are devoid of the pathological behavior shown in Fig.\ref{rank}.
            We will thus restrain our study to the
books with ranks below $10,000$.

    \subsection{Evidence that the rank-sales relation is close to a power law}
    \label{evidence conversion power law}

    We present the findings of M. Rosenthal \cite{surfing amazon}. For
    over six years, he has followed closely the sales ranks of his own
    books as both an author and a publisher. He has also used data points
    from other authors and publishers. Note that his analysis is
    unauthorized and in no way sponsored by Amazon, which keeps the sales-ranks
    conversion secret. His analysis allows us to get an approximate information
    on how to convert ranks into sales.

                     \begin{table}[htb]
                     \begin{center}
                     \begin{tabular}{|c|c|}
                       \hline
                       Rank & Copies/day \\
                       \hline
                       1 & $X$ \\
                       10 & 100 \\
                       100 & 30 \\
                       1000 & 10 \\
                       10,000 & 2  (11 copies every 5 days) \\
                       100,000 & 1 copy a week \\
                       1000,000 & around 15 total, depends on pub. date \\
                       \hline
                     \end{tabular}
                     \caption{Estimate of M. Rosenthal \cite{surfing amazon} on
                     the rank-sale relationship.}
                     \end{center}
                     \label{table1}
                     \end{table}

\begin{figure}
                \begin{center}
                  \includegraphics[width=0.7\textwidth]{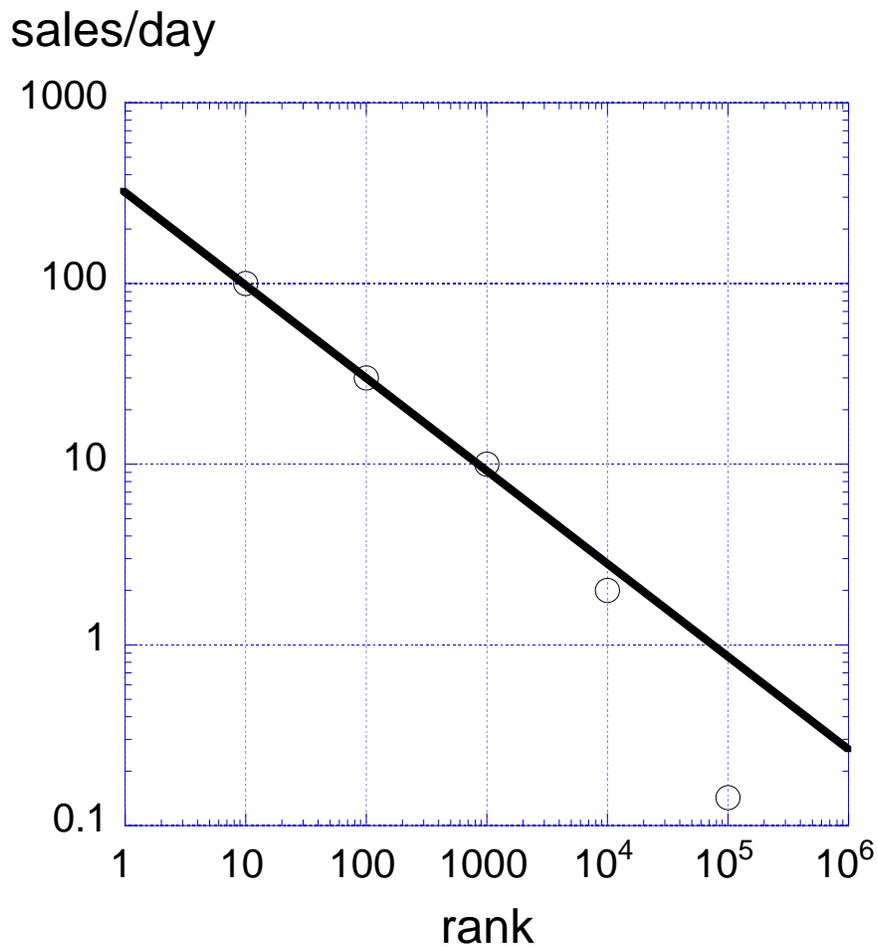}
                  \caption{Rank ordering (Zipf)
plot of the sales $S$ per day, as a function of
rank $R$.
                   For $R$ in the range from $10$
to $10,000$, the sales as a function of rank
                   can be fitted with a good approximation by the power law
                   $S(R) \sim 1/R^{1/\mu}$ with exponent $\mu = 2.0 \pm 0.1$, as
                   shown by the straight line with
slope $-0.5$. This translates into
                   $R \sim 1/S^{\mu}$, which is
proportional to the complementary cumulative
                   distribution of sales. Note
that the bend at large ranks (small sales)
                   can be considered to be a
finite size effect. The tail of the distribution
of sales for large sales is
described by the low ranks for which the
distribution does not appear to bend but
actually exhibits huge fluctuations as shown in figure
\protect\ref{rosenthal2} borrowed from
\protect\cite{surfing amazon}.}
        \label{rosenthal}
                \end{center}
    \end{figure}

Table I shows the estimates of M. Rosenthal, which are converted
into a rank-ordering or Zipf plot in Fig.~\ref{rosenthal}. The power
law dependence $S(R) \sim 1/R^{1/\mu}$ with exponent $\mu = 2.0 \pm
0.1$, as shown by the straight line with slope $-0.5$ in
Fig.~\ref{rosenthal} translates into a standard power law of the
complementary cumulative distribution of sales \be R \sim
1/S^{\mu}~,~~~~~~{\rm with}~~ \mu = 2.0 \pm 0.1~.
\label{pdfranksales} \ee The bend for large ranks (small sales) is
describing the bulk of the distribution of sales. The tail of the
distribution of sales, i.e., for large sales, is described by the
part of the figure corresponding to small ranks for which there does
not appear to be a change of regime from a power law behavior,
except for the fact that sales for the first ranks seem to
fluctuation from book to book much more significantly that would be
expected from a pure power law behavior. The current blockbuster
with rank $1$ may sell from hundreds of books per day to as much at
tens of thousands of books per day. Such variations may perhaps be
explained from the property that, while the typical fluctuations of
the sales of the first rank is of the order of a few 100\%
\cite{Sorfirstbook}, the distribution of the sales of the first rank
is also a power law distribution of the form (\ref{pdfranksales})
with the same exponent, which means that it is not impossible to
have very large variations of the sales of rank 1, much larger than
their typical values. These fluctuations are rendered in
Fig.~\ref{rosenthal2} in the left part for small ranks. More data
would be needed to determine if the variations of the sales of the
first ranks are explained by the power law distribution
(\ref{pdfranksales}) or may perhaps reveal an amplification
mechanism putting blockbusters apart. It is also possible that the
increased fluctuations for the smallest ranks reflect the effect of
the network structure of acquaintances. In contrast, the
fluctuations of sales from rank 10 are typically of the order of
30\% in agreement with expectations derived from the power law
(\ref{pdfranksales}).

The correspondence between ranks and sales
suggested in Figures \ref{rosenthal} and
\ref{rosenthal2}
and captured by the phenomenological formula (\ref{pdfranksales})
allows us to convert the time series of ranks for all studied books into
time series of sales. Notwithstanding the possible uncertainties and errors
in the calibration of the conversion from ranks
to sales, as discussed above, one
should consider this conversion as basically a
convenient way to give a quantitative
interpretation to the rank time series.

In summary, we will assume that the pdf $p(S)$ of
  book sales is a power law of the form
     \begin{equation}
      p(S)=\frac{C}{S^{1+\mu}},
      \label{pdf S}
      \end{equation}
      with $\mu=2$. In section \ref{Direct measure
on rank}, we revisit this assumption
      and test it further from constraints on  the averages of rank increments.

\section{Description of the Model}
\label{Description of the Model}

       \subsection{Motivation: exogenous versus endogenous shocks}

\begin{figure}
                \begin{center}

\includegraphics[width=.65\textwidth]{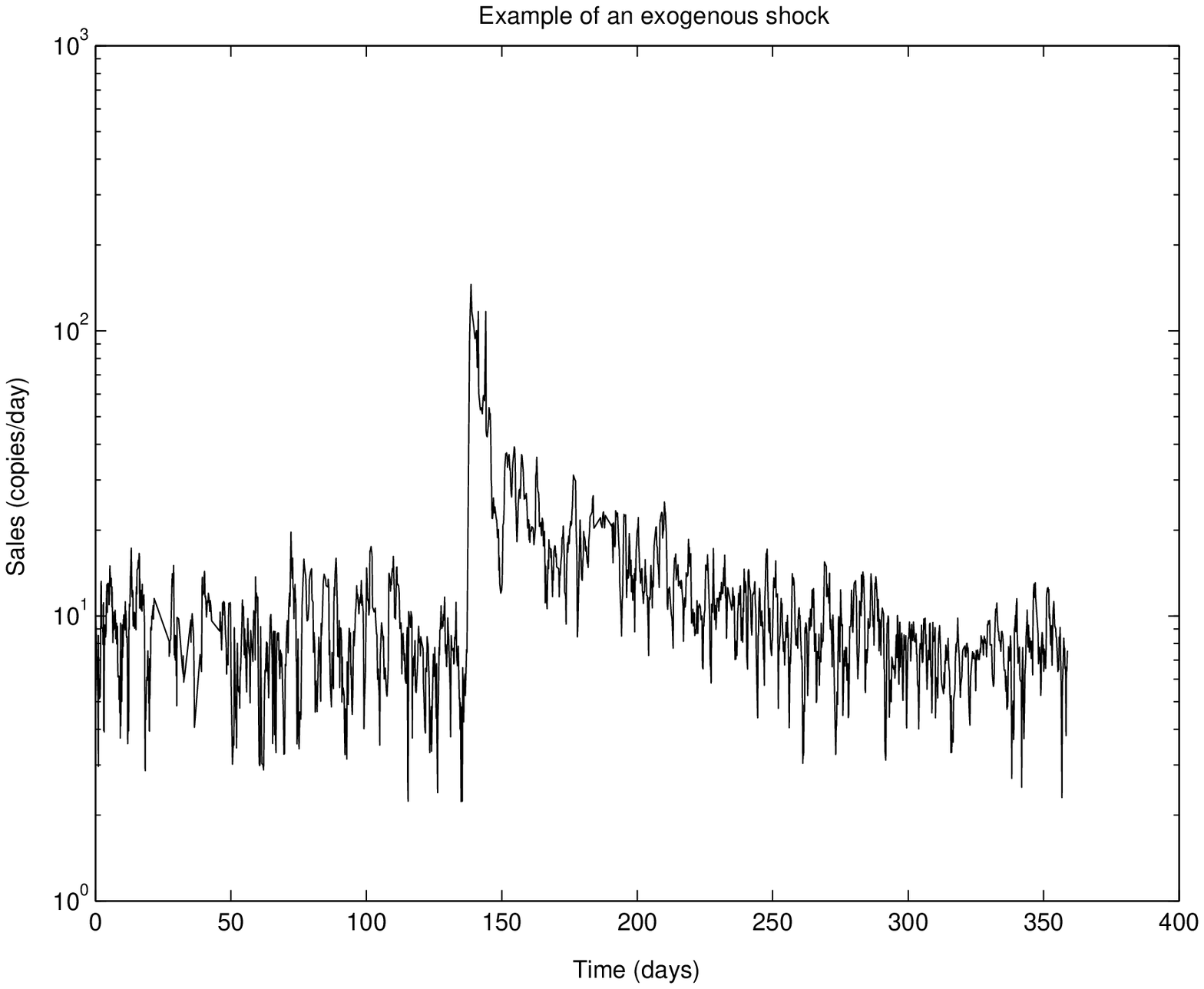}
                  \end{center}
                 \begin{center}

\includegraphics[width=.65\textwidth]{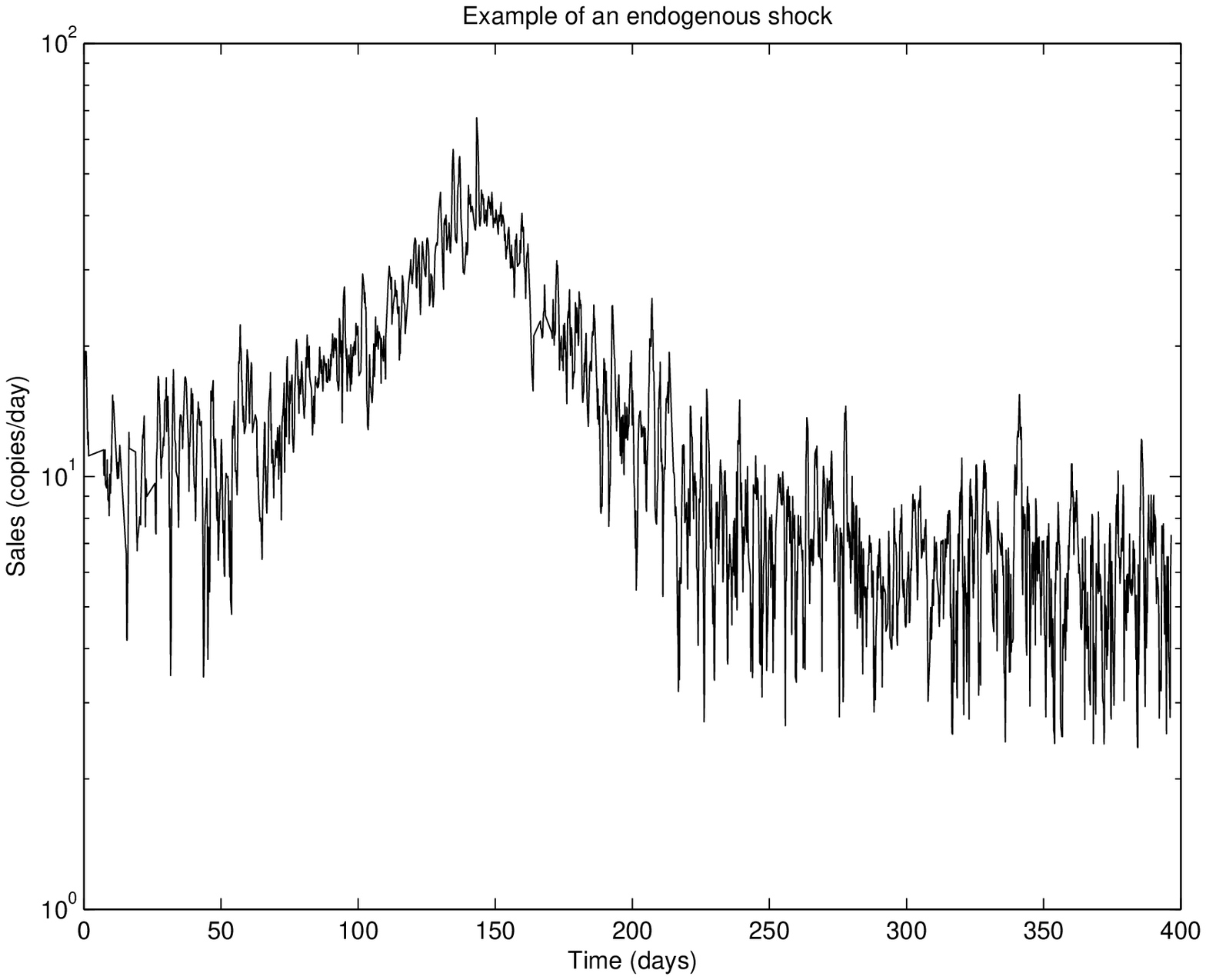}
                  \caption{Time evolution over a
year of the sales per day of two books :
                  Book A (top) is ``Strong Women Stay Young'' by Dr. M. Nelson
                and Book B (bottom) is ``Heaven
and Earth (Three Sisters Island Trilogy)''
                by N.Roberts. The difference in
the patterns is striking, Book A (resp. B)
                 exhibiting an exogenous (resp. endogenous) peak.}
                \label{sales}
                \end{center}
       \end{figure}

         Consider the two sales time series shown
in Fig.\ref{sales}, exemplifying
         two characteristic patterns.
         \begin{itemize}
         \item Some books become best-sellers overnight, thanks to rocketing
         sales. Book A (``Strong Women Stay Young'' by Dr. M. Nelson)
         jumped on June 5, 2002 from rank in the
2,000s to rank 6 in less than $12$ hours.
         On June 4, 2002, the New York Times
published an article crediting the
``groundbreaking
         research done by Dr. Miriam Nelson'' and advising the female
         reader, interested in having a youthful
postmenopausal body, to buy the book and
         consult it directly \cite{new york times}. This case is
         the archetype of what we will refer to as an ``exogenous'' shock.
         \item Some books become best-sellers
after a long and steady increase in their
         sales. Book B (``Heaven and Earth (Three
Sisters Trilogy)'' by N. Roberts)
         culminated at the end of June
         2002 after a slow and continuous growth, with no such newspaper
         article, followed by a similar almost symmetrical decrease, the
         entire process taking about $4$ months.
         \end{itemize}

       \subsection{Epidemic branching process with long-range memory}
       \label{Epidemic branching process}

       Such social epidemic process can be
captured by the following simple model
       \cite{system with memory}. The model is based on the idea that the
     instantaneous sales flux of a given book results from a
     combination of external forces such as news, advertisement,
     selling campaign, and of social influences in which each past
     reader may impregnate other potential readers in her network of
     acquaintances with the desire to buy the book. This impact of a
     reader onto other readers is not instantaneous as people react at
     a variety of time scales. The time delays capture
the time interval between social encounters, the maturation of the
decision process which can be influenced by mood, sentiments, and
many other factors and the availability and capacity to implement
the decision. We postulate that this latency can be described by a
memory kernel $\phi(t-t_i)$ giving the probability that a buy
initiated at time $t_i$ leads to another buy at a later time $t$ by
another person in direct contact with the first buying individual.
We consider the memory function $\phi(t-t_i)$ as a fundamental
macroscopic description of how long it takes for a human to be
triggered into action, following the interaction with an already
active human. $\phi(t)$
     is normalized such that $\int_0^{\infty}\phi(t)=1$.
     Starting from an initial
     buyer (the ``mother'' buyer) who notices the book (either from
     exogenous news or by chance), she may trigger buying by
     first-generation ``daughters,'' who themselves propagate the
     buying drive to their own friends, who become second-generation
     buyers, and so on. We describe the sum of all buys by a
     conditional Poisson branching process with intensity:
     \begin{equation}
     \label{branching process}
     \lambda(t) = \eta(t) + \sum_{i|t_i \leq t} \mu_i \, \phi(t-t_i)~,
     \end{equation}
     where $\eta(t)$ is the rate of buys initiated
spontaneously (for instance by
     listening to a media coverage of a book or serendipity)
     without influence from other previous buyers and the mark $\mu_i$ is
     the number of potential buyers influenced by the buyer $i$ who
     bought earlier at time $t_i$.

     Our model is based on the key idea that the epidemic mechanism
     is basically the same for all books. Of course, the underlying networks
     of connected potential buyers are a priori
not the same for different books.
     This can be accounted for by different values of the ``branching ratio''
     as defined below.

     While this version of the epidemic model of sales treats each
     book independently, in reality, we should consider correlations
     between sales of different books which may be related by a
     common growth of interest (see for instance the case of books
     on financial markets whose sales grow concomitantly during stock
     market bubbles \cite{RoehnerSorfrenzy}). In addition, sales
     will exhibit some correlation at special epochs, such as
     Christmas. At this period of the year, all
     books that have a gift appeal will sell more copies than they
     would have sold otherwise. In contrast,
     if a book does not have any gift appeal, its sales ranks will fall
     between Thanksgiving and Christmas, even if its actual sales remain
     steady. Likewise, university students buy a huge number of textbooks
     and other required reading titles through Amazon during September
     and from mid-January through mid-February, which will depress the
     ranks of books that do not fall into this category. They even vary in
     a regular manner during the course of the week. Some titles are
     primarily purchased by people at work or homemakers when the kids
     are at school, while books with strong Associates support do
     relatively well on weekends. We will not take into account such
     effects.

     We do not specifically describe the lifetime of a book and treat the
     innovations $\eta$'s as stationary. Assuming sales as stationary
     allows us to define the probability that the sales reach a
     given value. This approximation should not be too bad over the time
     scales of months of our study but fails over longer time scales.

     \subsection{Mean field solution}
     \label{Mean field equation}

     Equation (\ref{branching process}) can be written
     \begin{equation}
     \label{branching processad}
     \lambda(t) = \eta(t) + \int^t \int  N[d\tau \times d\mu] \phi(t-\tau)~
        \end{equation}
     where $N[d\tau \times d\mu]$ is the standard notation for the number of
     events that occurred between $\tau$ and $\tau + d\tau$ with mark between
      $\mu$ and $\mu + d\mu$.  In the physicist's notation
      $ \int^t \int  N[d\tau \times d\mu] =
\int^t \int  \delta(\tau -
t_i)~\delta(\mu-\mu_i)$.
      The lower bound of the integral over time in
(\ref{branching processad}) is for instance
      the edition time of the considered book.
      Taking the ensemble average of (\ref{branching process}) gives
      \begin{equation}
      \label{mean field}
      S(t) \equiv \langle \lambda(t) \rangle = \eta(t) + n \int_{-\infty}^{t}
      \mathrm{d}\tau ~\phi(t-\tau) S(\tau),
      \end{equation}
      where the so-called ``branching ratio''
      $n=\langle \mu \rangle$ is the average number of buys triggered by
      any ``mother'' within her acquaintance network. We have use the fact that
      $\langle \int  N[d\tau \times d\mu] \rangle
= n \langle \lambda(\tau) \rangle$.
      The branching ratio $n$ depends on
      the network topology as well on the social behavior of
      influences. We consider only the
sub-critical regime $n<1$ in order to ensure
stationarity.
       The linear structure of equation (\ref{mean field}) does not mean
       that the dynamic is  linear. It is an effective coarse-grained
       description of the complex nonlinear dynamics.

      In order to solve $S(t)$, it is convenient
to introduce the Green function or
      ``dressed propagator'' $\kappa(t)$ defined
      as the solution of (\ref{mean field})  for the case where the source term
      $\eta(t)$ is a delta function centered
      at the origin of time :
      \begin{equation}
      \kappa(t) = \delta(t) + n \int_{-\infty}^{t}
\mathrm{d}\tau ~\phi(t-\tau)\kappa(\tau),
      \label{equation sur kappa}
      \end{equation}
       and by definition of $\kappa(t)$ :
      \begin{equation}
      S(t)=\int_{-\infty}^{t} d\tau ~\eta(\tau)~ \kappa(t-\tau).
       \label{linear equation}
       \end{equation}
       The cumulative effect of all the possible
branching paths gives rise to the net
       sales flux $\kappa(t)$ triggered by the initial event at time $t=0$.
         The response function $\kappa(t)$
       can easily be obtained by taking
      the Laplace transform of (\ref{equation sur kappa}):
      \begin{equation}
      \widehat{\kappa}(\beta) = \frac{1}{1-n\widehat{\phi}(\beta)}
      \end{equation}
     Setting $\beta=0$, we get
       \begin{equation}
       \int_0^{\infty}\mathrm{d}\tau \, \kappa(\tau)=\frac{1}{1-n}~,
       \end{equation}
       which means that  $\frac{1}{1-n}$ is the
average number of buyers influenced
        by one buyer through any possible lineage. This result can be recovered
        directly from the following argument: if
$n$ is the average  number of buyers influenced
        directly by one buyer, the total number of buyers influenced
        through any possible lineage is $\sum_{k=0}^{\infty} n^k=
        \frac{1}{1-n}$. The term $\frac{\kappa(t)}{1-n}\mathrm{d}t$ is
        the probability that a purchase
triggered by a buy at $t=0$ occurs at time t within $\mathrm{d}t$.

      We consider the case where the ``bare
propagator'' is $\phi(t) \sim 1/t^{1+\theta} $
with
      $0<\theta<1$ corresponding to a long memory process. It leads to :
      \begin{align}
      &\kappa(t) \sim 1/t^{1-\theta},
~~~~\mathrm{for} \, \, \, t<t^\star
\label{kappa1}\\
      &\kappa(t) \sim 1/t^{1+\theta}, ~~~~\mathrm{for} \, \, \, t>t^\star,
      \end{align}
      with
      \begin{equation}
      t^\star \propto 1/(1-n)^{1/\theta}.
      \label{t star}
      \end{equation}

\subsection{Prediction of the model}

\subsubsection{Distribution of sales}

      Starting from the evidence that the distribution $p(S)$ of sales
      is a power law (\ref{pdf S}), the linear
expression (\ref{linear equation})
      implies that the source terms $\eta(\tau)$
in (\ref{linear equation}) are also
      distributed as a power law with the same exponent $\mu$. Actually,
      the generalized central limit theorem applied to (\ref{linear equation})
      implies that the pdf $p(S)$ is a stable
L\'evy law $L_{\mu}$ with index $\mu$
      as soon as the source terms $\eta(\tau)$ in (\ref{linear equation}) are
      independently and identically
      distributed as a power law with an exponent $\mu \leq 2$. By the
      generalized central limit theorem, the characteristic
      function of $S$ is given by
      \be
      \langle e^{iuS(t)} \rangle = \exp \left[ -D|u|^{\mu}~\int_0^{\infty}
       |\kappa(\tau)|^{\mu}~d\tau \right]~,
      \label{ngnjsdks}
      \ee
      where $D$ is a measure of the magnitude of the sources $\eta(\tau)$.
      This translates into
      \be
      P(S) \approx L_{\mu}\left[{S \over \left(D \int_0^{\infty}
       |\kappa(\tau)|^{\mu}~d\tau \right)^{1 /\mu} }\right]~.
       \ee
      In reality, there is probably a dependence
between the source terms $\eta(\tau)$.
      However, as long as their correlation function decays faster than $1/t$,
      this has only the effect of changing the coefficient $D$. We do not
      explore the situation in which the $\eta(\tau)$'s could have longer range
      correlations.

\subsubsection{Dynamical properties}
\label{dynamical properties}

                \paragraph{Exogenous shock.}
                  An external shock occurring at $t=0$ can be modeled as a jump
                  $S_{0}\delta(\tau)$.
                  The response of the system for $t>0$ is then :
                  \begin{equation*}
                  S(t)=S_{0}~ \kappa(t) +
\int_{-\infty}^{t} \eta(\tau)\kappa(t-\tau).
                  \end{equation*}
The expectation of the response to an exogenous shock is thus :
                  \begin{equation}
\mathrm{E}[S(t)]=S_{0}~\kappa(t)+\frac{1}{1-n} \langle \eta \rangle,
                   \label{E exo}
                 \end{equation}
                 where $\langle \eta \rangle$ is the average source level.
                 Expression (\ref{E exo})
                 simply expresses that the recovery of the system
                 to an external shock is entirely controlled by its
                 relaxation kernel.

                 For an external shock which is strong enough, for $t>0$ :
                 \begin{equation}
                 \mathrm{E_{exo}}(S(t)) \approx S_0 ~ \kappa(t).
                 \label{E_exo}
                 \end{equation}
                It exemplifies that $\kappa(t)$ is the
                 Green function of the coarse-grained equation of
                 motions of the system.

                \paragraph{Endogenous shock.}

                 Consider a realization which exhibits a large sales burst
                 $S(t=0)=S_{0}$ without any large
external shock. In this case, a large
                 endogenous shock requires a special set of
                 realizations of the noise $\{\eta(t)\}$
                 \cite{system with memory}. We can
write : $\eta= \widetilde{\eta}\, +
                 \langle \eta \rangle$. By
construction  $\widetilde{\eta}$ has a zero
                 mean. Using this, we get :
                 \begin{align*}
                 S(t)=
                 &\int_{-\infty}^{0} \mathrm{d}\tau
                 \widetilde{\eta}(\tau)\kappa(t-\tau) +
                 \int_{-\infty}^{0}\mathrm{d}\tau
\langle \eta \rangle \kappa(t-\tau)
                 \nonumber\\ + &\int_{0}^{t}\mathrm{d}\tau \eta(\tau)
                 \kappa(t-\tau)
                 \end{align*}
                 The expectation of $S$ is :
                 \begin{equation}
                 \mathrm{E}[S(t)]=\int_{-\infty}^{0} \mathrm{d}\tau
                 \mathrm{E}\left[\widetilde{\eta}(\tau)\right]\kappa(t-\tau) +
                  \frac{\langle \eta \rangle}{1-n}~.
                 \label{expectation sales endogenous}
                 \end{equation}
                  As for an exogenous shock, the constant
                  $\frac{\langle \eta
\rangle}{1-n}$, equal to the unconditional average
                  $\langle S \rangle$, can be neglected.

                  For $\tau<0$, the expectation of
                  $\widetilde{\eta}(\tau)$ is not zero, because
                  the value $S(0)=S_0$ is specified. In contrast,
                  for $\tau>0$,
$\mathrm{E}[\eta(\tau)]= \langle \eta \rangle$
since
                  the conditioning does not operate after the shock.
                  Consider the process $W(t) \equiv
                  \int_{-\infty}^t \mathrm{d}\tau~
                  \widetilde{\eta}(\tau)$. A standard result is that for t<0:
                  \begin{align*}
\mathrm{E}\biggl[W(t)|S(0)=S_{0}\biggr]&= (S_{0}-\mathrm{E}[S]) .
                  \frac{\mathrm{Cov}[W(t),S_{0}]}{\mathrm{Var}[S_{0}]}
                   \\ & \simeq
                  (S_{0}-\mathrm{E}[S]) \int_{-\infty}^{t} \mathrm{d}\tau
                  \, \kappa(-\tau) ~.
                  \end{align*}
                 This expression predicts that the expected path of the
                  continuous innovation flow prior to the endogenous
                  shock (i.e. for $t<0$) grows like $\Delta
                  W(t) \sim \kappa(-t) \Delta
                  t$ upon the approach to the time $t=0$ of the large
                  endogenous shock. In other words, conditioned on the
                  observation of a large endogenous shock, there are
                  specific sets of the innovation flow that led to it.
                  These conditional innovation flows have an
                  expectation
$\mathrm{E}[\eta(t<0)]- \langle \eta \rangle
\simeq
                  S_{0} \kappa(-t)$ (We assume $S_{0} \gg \mathrm{E}[S]$).
                  We thus obtain from (\ref{expectation sales endogenous})
                  for $t>0$ and $t<0$ :
                  \begin{equation}
                  \mathrm{E_{endo}}[S(t)] \propto S_{0}
                  \int_{-\infty}^{{\rm Min}[t,0]}
\mathrm{d}\tau \, \kappa(t-\tau) ~
                  \kappa(\tau) ~.
                  \label{E_endo}
                  \end{equation}

                \paragraph{Distinguishing both shocks}
                 The model predicts two different relaxations for
                 exogenous and endogenous shocks
according to expressions (\ref{E_exo}) and
                 (\ref{E_endo}). Assuming that we are close to the
                 critical point $n \approx 1$, we can use
                 $\kappa(\tau) \sim 1/t^{1-\theta}$. Table
                 \ref{distinguishing shock} gathers the aftershock and
                 foreshock signatures.

                 \begin{table}[h]
                     \begin{center}
                     \begin{tabular}{|c|c|c|}
                       \hline
                        & Endo & Exo \\
                       \hline
                       Aftershock & $S(t) \propto
\frac{1}{t^{1-2\theta}}$& $S(t)
                       \propto \frac{1}{t^{1-\theta}}$\\
                       \hline
                       Foreshock & $S(t) \propto
\frac{1}{|t|^{1-2\theta}}$& Abrupt Peak\\
                       \hline
                     \end{tabular}
                     \caption{Aftershock and
foreshock signatures for endogenous and exogenous
                     shocks occurring at time $t=0$.}
                     \label{distinguishing shock}
                     \end{center}
                     \end{table}

                     The prediction that the
relaxation following an exogenous shock
                       should happen faster
(exponent $1-\theta$) than for an endogenous shock
                     (exponent $1-2\theta$) agrees
with the intuition that an endogenous shock
                     should have impregnated the
network much more and thus have a longer
                     lived influence. This result
is a non-trivial consequence of our model. If we
                     perform the same calculation
\cite{system with memory} for an
                     exponential decaying memory kernel $\phi(t)$,
                     the functional form of the recovery does not
                     allow one to distinguish between an endogenous
                     and an exogenous shock.
                     For the memory kernel $\phi(t)$ decaying
                     faster than an exponential, the endogenous
                     relaxation turns out to be
faster than the exogenous one \cite{system with
memory}.

\section{Empirical determination of the exponents of sales dynamics}
\label{sdth}

   \subsection{Selection of peaks and fitting the power law}
   \label{Determining the exponents}

We qualify a peak as a local maximum over a $3$-month time window which
is at least $k=2.5$ time larger than the average of the time series over
the same $3$-month time window. The threshold
value $k$ was determined after looking at
several examples of time series. The results do not change significantly by
varying $k$ within large bounds (see below).

We considered only the sales maxima corresponding
to ranks reaching the top $50$.

We selected those time series which had at least $15$ days after the peak, so
that we can analyze the recovery signature following shocks.

         We fit the sales dynamics by a power law :
         \begin{equation*}
            S(t) \sim \frac{A}{(t-t_c)^p},
         \end{equation*}
         with $A$, $p$ and $t_c$ unknown. The
``critical time'' $t_c$ is expected to be close to
         the time $t_{0}$ of the peak. We know from previous experience in
         critical phenomena that the determination
of the exponent $p$ can be quite
         sensitive to the fitted value of $t_c$.

         Moreover, we do not know a priori which
window size should be taken. If the
         signal were a pure power law, it would not matter. But, here the
         signal is approximately a power law over a finite range. A big
         time window means more data and so more accurate results but we
         have to end the window before a possible change of regime due to
         (i) $n<1$ implying the existence of the
finite cross-over time $t^\star$
         given by (\ref{t star}), (ii) $S(t)$ tending to background noise around
         to its average value and (iii) the impact of other shocks that may
         interfere.

\begin{figure}
                \begin{center}
                  \includegraphics[width=0.7\textwidth]{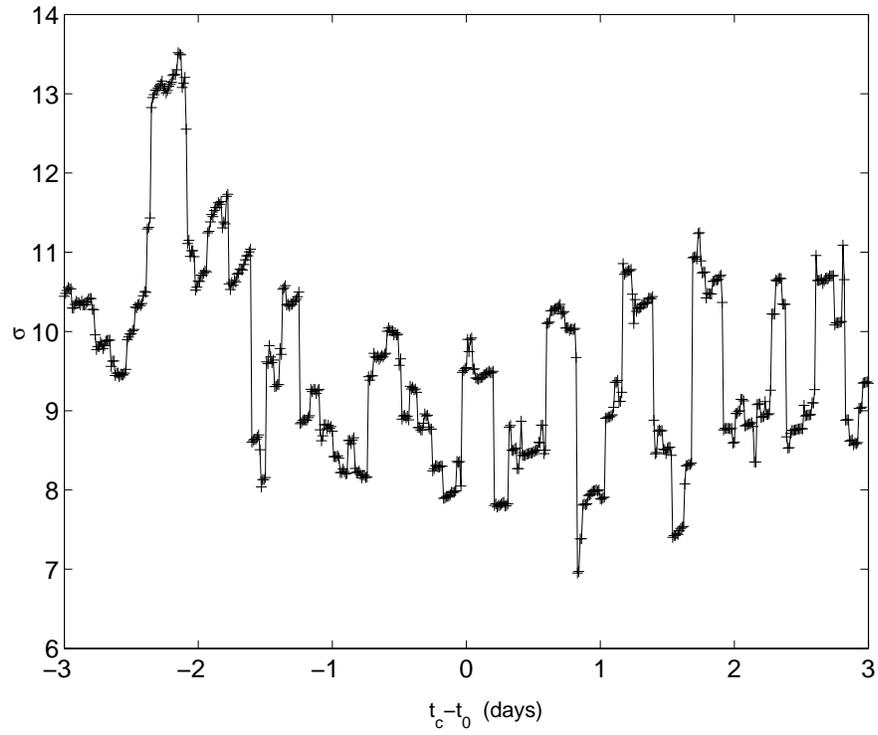}
                  \caption{$\sigma(t_c)$ defined
in (\ref{gmnmdlsd}) as a function of $t_c$.}
                  \label{landscape}
                \end{center}
\end{figure}

         In order to determine $A$, $p$ and $t_c$ for a given window
         size, we use a least square
         method, i.e. we seek to minimize the following quantity :
         \begin{equation*}
          \sigma_{log{A},p,t_c}  = \sum _{t_i} F_{log{A},p,t_c}(t_i)^2,
          \end{equation*}
        with,
        \begin{equation*}
        F_{log{A},p,t_c}(t_i) =  \log{S(t_i)} -\log{A} +p \, \log(t_i-t_c).
        \end{equation*}
        As $\sigma$ is quadratic with respect to
        $\log{A}$ and $p$, for these two
parameters, the minimization of $\sigma$ is
        straightforward and can be done analytically. Setting the
        partial derivatives $\frac{\partial \sigma}{\partial \log{A}}$ and
        $\frac{\partial \sigma}{\partial p}$  equal to zero, we
        just need to solve a linear system of two
equations with two unknowns, which
        gives $\log {A(t_c)}$ and $p_{t_c}$ as a
function of the still unknown $t_c$.
        Now, we just need to minimize
  \be
  \sigma_{t_c} = \sigma_{log{A(t_c),p(t_c),t_c}}~,
  \label{gmnmdlsd}
  \ee
          a  function of one variable.
         Fig. \ref{landscape} shows an example of
        $\sigma(t_c)$. To minimize such irregular function, we
        scan over all the value of $t_c$. We typically
        use a total interval of one week around the time of the peak.

  We perform a fit for different time windows ranging from a lower bound
  $I_{min}$ to an upper bound $I_{max}$. $I_{min}$ has a fixed value, set
  to $15$ days. In contrast, $I_{max}$ depends on the time series.
  $I_{max}$ is calculated as the time at which the minimum of the sales
  occurs, over a time window running from $25$ days to $6$ months.
  $I_{max}$ is fixed in this way to prevent a rise
  in sales from being taken into account in the fit.

  Once the fit has been performed for the different time windows, we look
  at the correlation coefficients of the fits and choose the
  window which leads to the best correlation coefficient.
  This method turns out to give a robust estimate
of the power law decay of the relaxation of
   sales.

\subsection{Distribution of exponents}
\label{Distribution of exponents}

Out of some $14,000$ books available on Junglescan on April 2004, our
algorithm detects $1,013$ peaks which obey the constraints
of reaching the top $50$, with sufficient data
before and after the peak and obeying the
condition of not being contaminated by a closeby peak, as specified above.
Among these $1,013$ peaks, we select those followed by a relaxation which
can be well-approximated by a power law, with the criterion that the
correlation coefficient $r$ of the corresponding fit is larger than $0.95$.
This leads us to keep $138$ peaks.
Making this selection does not change qualitatively our results but improve
somewhat the quantitative findings. We have played with different values of
the correlation coefficient between $0.8$ and close to $1$ and find
the same results with larger error bars for lower correlation coefficients.

\begin{figure}
                \begin{center}
                  \includegraphics[width=1\textwidth]{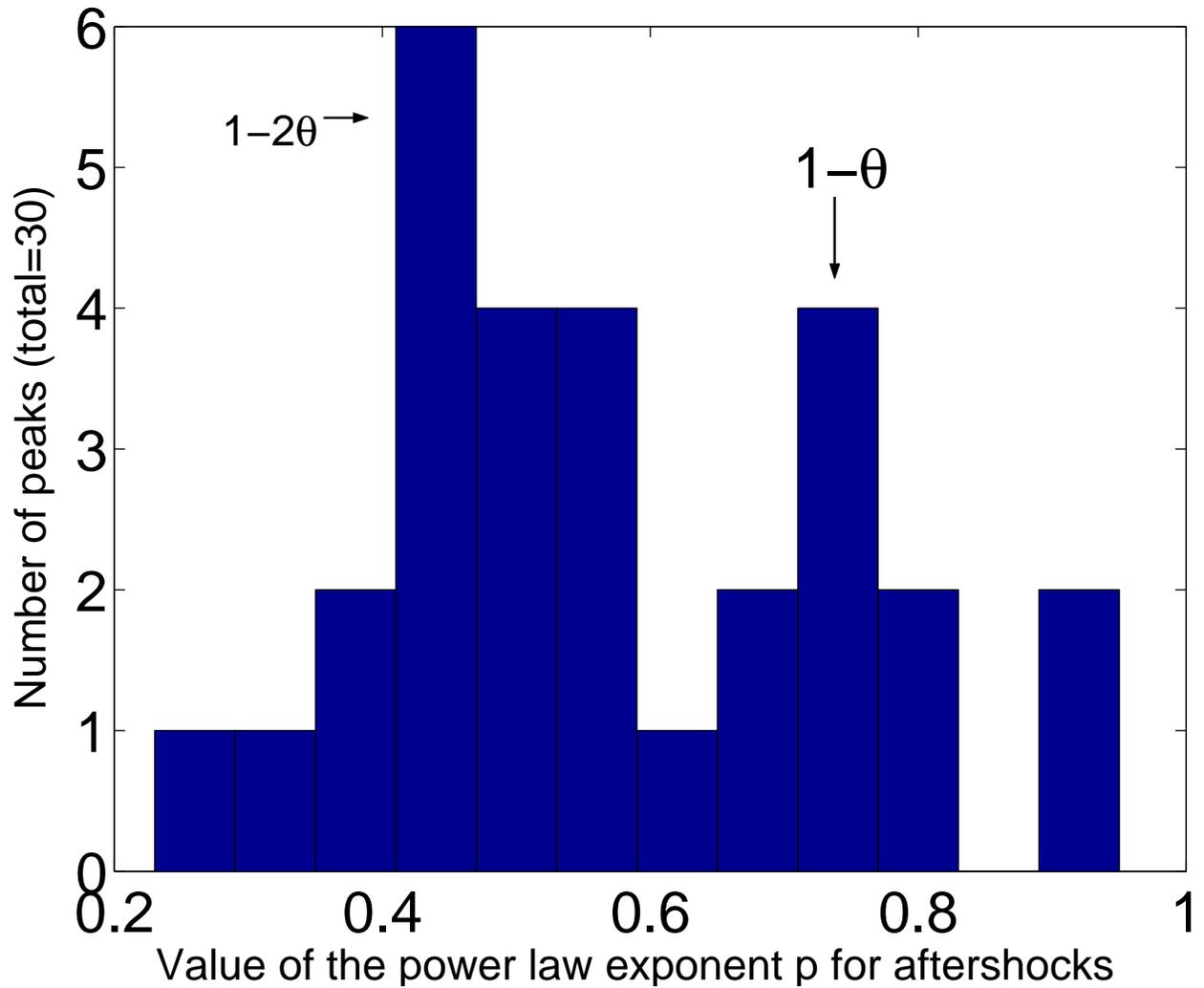}
                  \caption{Histogram of the estimated power law
                  exponents $p$ of the relaxations of the sales. One can
                  clearly identify two clusters: the endogenous
                  cluster with exponent
$1-2\theta$ close to $0.4$ and the exogenous
                  cluster with exponent $1-\theta$
close to $0.7$, compatible with
                  the estimation $\theta \approx
0.3$. The peaks shown here are those used
                  in
Fig.~\protect\ref{stack_after} (see section
\protect\ref{Stacking the peaks}).}
                   \label{histo}
                   \end{center}
\end{figure}

Fig.\ref{histo} shows the distribution of power law
exponents for the decrease of sales after peaks.
We find two clusters corresponding to peaks
respectively with an exponent $0.2<p<0.6$ and with an exponent $0.6<p<1$. This
suggests that these two clusters can be seen as the endogenous
cluster ($1-2\theta\approx 0.4$) and the exogenous one
($1-\theta\approx0.7$). This provides a first estimate for the
exponent $\theta \approx 0.3$.

According to the epidemic model proposed here, the small values of
the exponents (close to $1-\theta$ and $1-2\theta$) for the
exogenous and endogenous relaxations, respectively, imply that the
sales dynamics is dominated by cascades involving high-order
generations rather than by interactions stopping after
first-generation buy triggering. Indeed, if buys were initiated
mostly by the direct effects of news and advertisements without
amplification by triggering cascades in the acquaintance network,
the cascade model would predict an exponent $1+\theta$ given by the
``bare'' memory kernel $\phi(t)$. The values smaller than $1$ for
the two exponents for exogenous and endogenous shocks imply
accordingly that the average number $n$ (the average branching ratio
in the language of branching models) of impregnated buyers per
initial buyer in the social epidemic model is on average very close
to its critical value $1$, because the renormalization from
$\phi(t)$ to $\kappa(t)$ given by (\ref{kappa1}) only operates close
to the criticality characterized by the occurrence of large cascades
of buys. Reciprocally, a value of the exponent $p$ larger than $1$
would suggest that the associated social network is far from
critical.

\subsection{Stacking the peaks}
\label{Stacking the peaks}

       According to our model and as summarized in
table \ref{distinguishing shock},
       the peaks belonging to the cluster with
       high $p$ ($p \approx 0.7$) should be in the exogenous class, and
       therefore should be reached by the occurrence of abrupt jumps
       without detectable precursory growth. Alternatively, the peaks
       belonging to the cluster with $p \approx 0.4$ should be in the
       endogenous class, and therefore should be associated with a
       progressive precursory power law growth $1/(t_{c}-t)^p$ with
       exponent $p=1-2 \theta$.

       To check this prediction, the following algorithm categorizes the
       growth of sales before each of the peaks according to its
       acceleration pattern. We differentiate between peaks that have an
       increase in sales by a factor of at least $k_{\mathrm{exo}}$
       prior to the peak and peaks that have an increase in
       sales by a factor of less than $k_{\mathrm{endo}}$ at the same
       time. More precisely, we compared the value of
       the sales at the time of the peak (D day) and the average value of
       the sales from day D-4 to D-1.
       We tried several values for these two
coefficients $k_{\mathrm{exo}}$ and
$k_{\mathrm{endo}}$.

       We find that the bigger $k_{\mathrm{exo}}$, the largest is the
       exponent of the average relaxation for books that have an increase
       in sales by a factor more than $k_{\mathrm{exo}}$. Conversely,  we
       find that the smaller $k_{\mathrm{endo}}$, the smaller is the
       exponent of the average relaxation for books that have an increase
       in sales by a factor less than $k_{\mathrm{exo}}$. Both results
       agree with the intuition that, for selective criteria, we only
       keep peaks that are very easy to classify and thus reduce the
       probability to make a misclassification.

\begin{figure}
                \begin{center}
                  \includegraphics[width=1\textwidth]{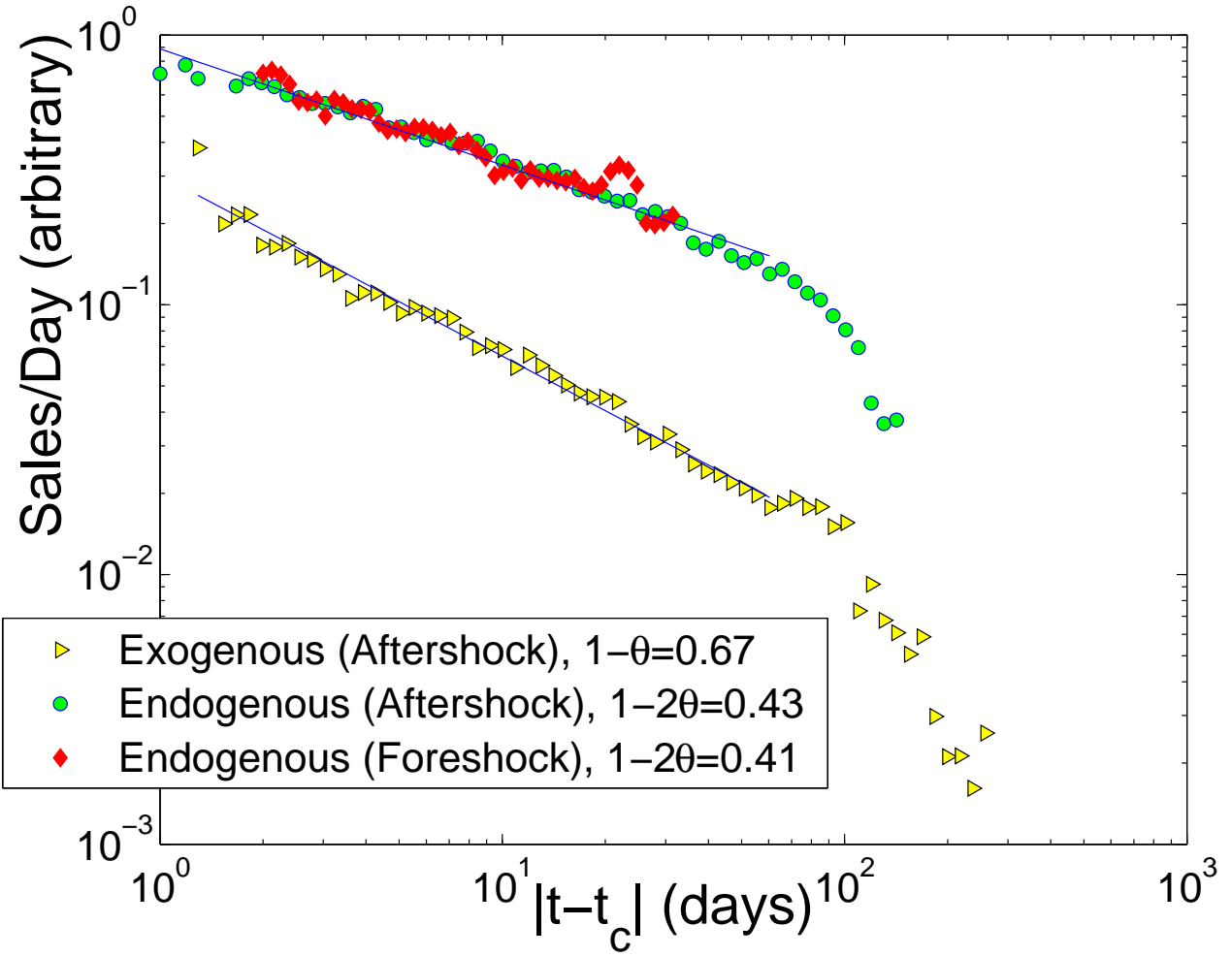}
                  \caption{Relaxation after the peak (for both endogenous
                  and exogenous cases) and precursory acceleration (for
                  the endogenous case). We average (log average) over the
                  peaks classified as endogenous and exogenous according
                  to their precursory growth. The
same sample of books was used for
                  both Fig.\protect\ref{histo} and
Fig.\protect\ref{stack_after}.}
  \label{stack_after}
\end{center}
   \end{figure}

       Finally, we set $k_{\mathrm{exo}} = 30$, $k_{\mathrm{endo}} = 2$.
       It means that peaks for which the acceleration factor were between
       $2$ and $30$, were not considered for the subsequent analysis
       leading to Fig.~\ref{stack_after}. Out of the $138$ peaks, $30$ remains.

      Fig.\ref{stack_after} shows the average relaxation and
       precursory acceleration. For shocks classified as exogenous
       according to their acceleration pattern, we find a relaxation
       governed by an exponent $1-\theta\approx0.7$. For shocks
       classified as endogenous, both aftershocks and foreshocks are
       controlled by an exponent $1-2\theta\approx0.4$. These results
       match what has been predicted by the model (see
       Table \ref{distinguishing shock}).

        \textbf{Aftershock:} the best fit of the
least square method with a power law gives a slope
$1-\theta \approx 0.7$ for exogenous shocks and a slope $1-2\theta
\approx 0.4$ for endogenous shocks. One can observe a crossover for
$t-t_c\approx 60-80\,\, \mathrm{days}$, in both cases. It is tempting to
interpret this crossover as the change of regime predicted by the model for
$t\approx t^\star$ (see section \ref{Mean field equation}). Indeed for
$t>t^\star$, we should expect a power law with exponent $1+\theta$, both
for exogenous and endogenous shocks. Unfortunately, the crossover does not
extend sufficiently far to allow us to constrain
the exponent of the second regime.

\textbf{Foreshock:} the best fit of the least square method with a
power law gives a slope $1-\theta \approx 0.4$ for the endogenous
foreshocks. The time on the $x$-axis has been reversed to compare the
precursory acceleration with the aftershock relaxation. The
superposition of the two top curves for the precursory and relaxation
behavior of the endogenous peaks confirms the symmetric behavior
predicted by the model (see Table \ref{distinguishing shock}).

\subsection{Detailed analysis of exogenous peaks}

\begin{figure}
    \begin{center}
      \includegraphics[width=0.7\textwidth]{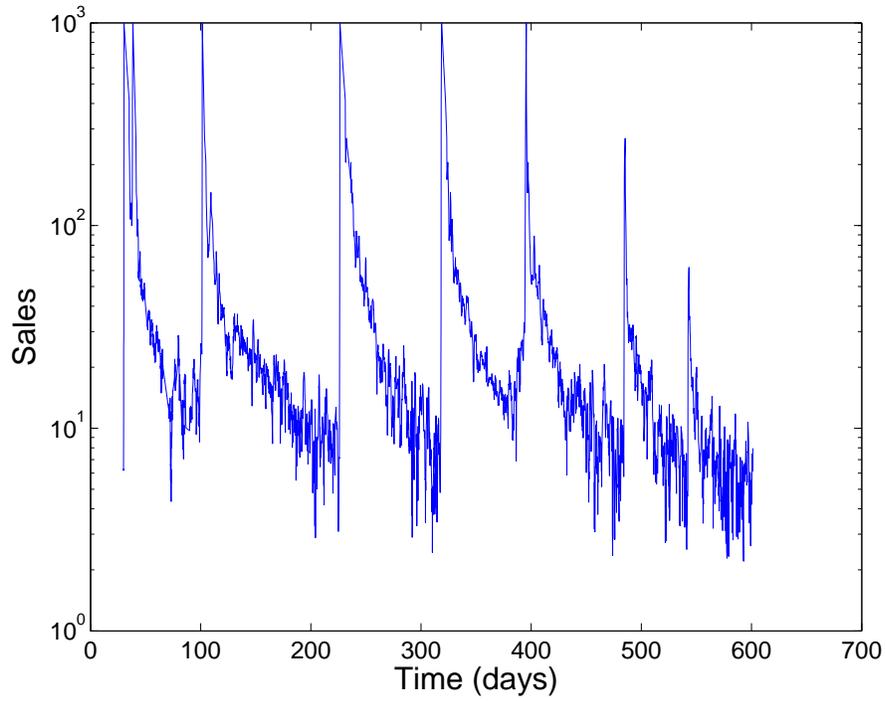}
      \caption{Time evolution of the book ``Get
with the Program'' (see Table \ref{list_peak})".
      Each time the book is presented in Oprah
Winfrey Show, the sales jumps overnight and
      then relaxes according to the exogenous
response function $\kappa(t) \sim
1/t^{1-\theta}$.}
      \label{oprah10}
    \end{center}
\end{figure}

\begin{figure}
    \begin{center}
      \includegraphics[width=1\textwidth]{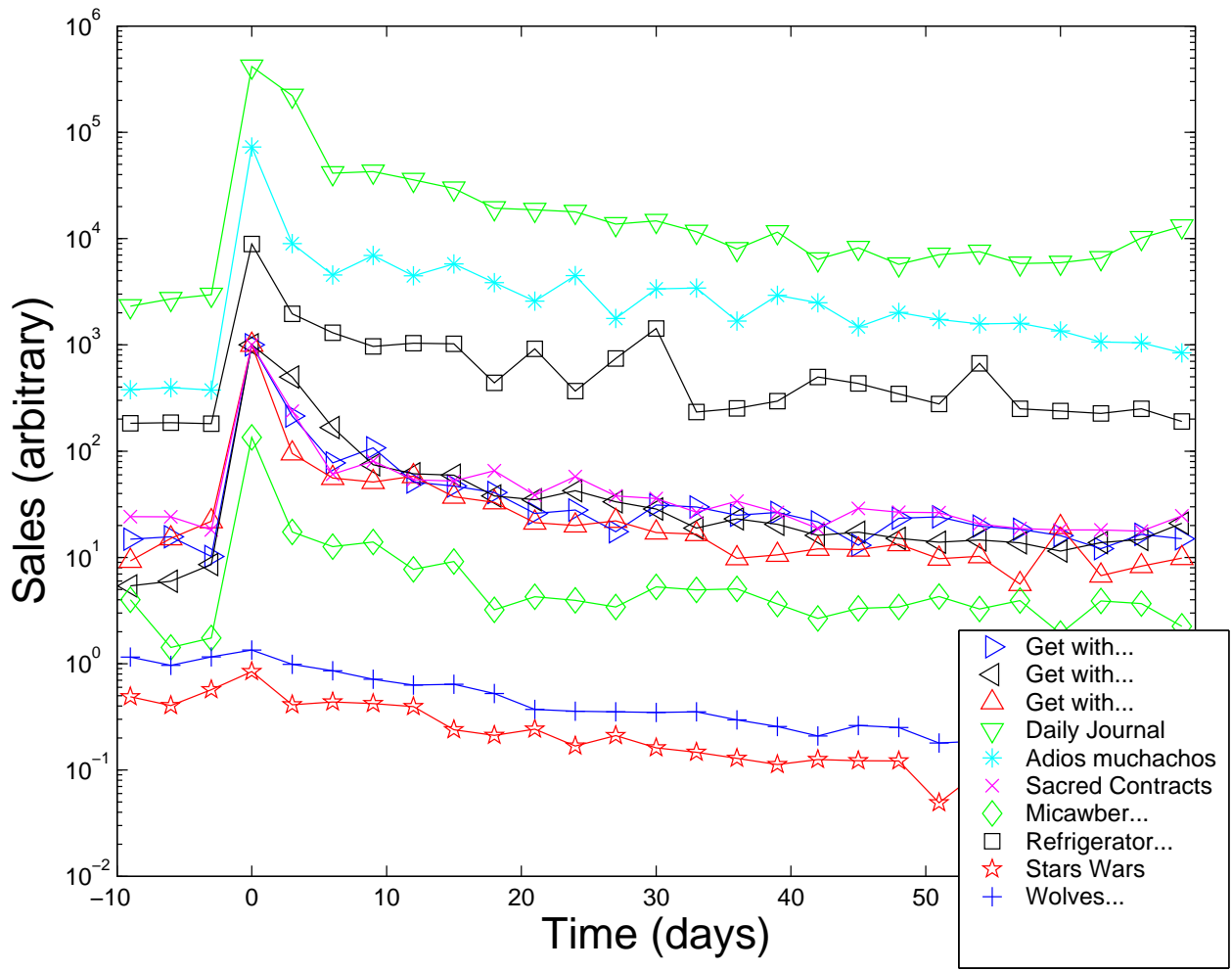}
      \caption{Time evolution of the sales of the
10 books of Table \ref{list_peak}.
      For each time series, $t=0$ is the
      time of the peak. The sales jumped just
before the peak except for ``Stars Wars'' and
      ``Wolves of the Calla.''}
      \label{10peaks}
    \end{center}
\end{figure}

Table \ref{list_peak} lists the $10$ peaks that have an exponent
larger than $0.65$ among the thirty peaks used to make
Fig.~\ref{histo}. Fig.~\ref{10peaks} shows the evolution of the
sales for these $10$ peaks, from ten days before to 70 days after
the peaks. Fig.~\ref{oprah10} shows the time evolution of the sales
of the book ``Get with the program'' written by the personal trainer
of Ophra Winfrey and its remarkable succession of exogenous peaks
associated with regular appearance of the book in Oprah's TV show.

In Fig.~\ref{10peaks}, eight peaks are reached by a fast
acceleration of the sales, as expected from our model and the
classification. However for two books, ``Star Wars'' and Stephen
King's novel, the situation is different. Their acceleration
patterns  are classified as endogenous (i.e. slow acceleration
growth) by our algorithm. Yet, they exhibit a fast relaxation (large
exponent $p=1-\theta$) which means that they are classified as
exogenous according to the first criterion, based on the exponent of
the relaxation after the peak.  So, what's wrong ?

   \begin{table}[htb]
                     \begin{center}

\begin{tabular}{|p{5cm}|p{2.5cm}|p{1.8cm}|p{.7cm}|p{5cm}|p{10.5cm}|}
                       \hline
                       Title & Author & Time & Rank &  Reason  \\
                       \hline
                       Get with the Program & Bob
Greene &  3/16/2002 & 1st & The book was seen on
the Oprah Winfrey Show (B.Greene is O. Winter's
trainer).  \\
                                ''          &
''      &  10/19/2002& 1st &     ''
\\
                                 ''           &
''      &  1/4/2003  & 1st &    ''
\\
                       Get with the Program Daily
Journal & Bob Greene & 10/19/2002& 2nd &  The
book was seen on the Oprah Winfrey Show (B.Greene
is O. Winter's trainer).\\
                       Adios Muchachos & Daniel
Chavarr\'ia & 7/5/2002 & 6th & On 7/4/2002, the
book was on the first page of the Art Section of
the New York Times\\
                       Sacred Contracts :
Awakening Your Divine Potential & Caroline Myss &
8/16/2002&1st& ?  \\
                       Micawber's Museum of Art &
John Lithgow & 9/30/2002 & 3rd & ?  \\
                       Refrigerator Rights :
Creating Connections and Restoring Relationships&Dr.Will
Miller & 1/26/2003 & 12nd &  The author has
appeared on a variety of
                                    national television programs\\
                       Stars Wars : Episode II,
Attack of the Clones & RA Salvatore & 5/20/2002 &
13rd & The movie was released on 5/16/2002. \\
                       Wolves of the Calla (The
Dark Tower, Book 5) & Stephen King& 11/11/2003 &
3rd & On 11/19, the 2003 Medal for Distinguished
Contribution to American
                           Letters was conferred upon Stephen King.\\

                      \hline
                     \end{tabular}
                     \caption{List of the 10 peaks
that have an exponent $p$ more than 0.65
                     (see
\textbf{Fig.\ref{histo}}). The last column
suggests a possible cause
                     of the exogenous shock, as
far as we have been able to tell.
                     }
                     \label{list_peak}
                     \end{center}
\end{table}

Consider the example of the book entitled ``Stars
Wars.'' Its success was triggered by the
release of the movie. But, as the advertisement campaign lasted several
weeks, not to say several months, no wonder we don't  observed a jump
but a slow acceleration growth. For such huge selling campaign, modeling
the source term $\eta(t)$ responsible for the
shock by a Dirac function is a very poor
approximation.
Here, the time duration of the advertisement campaign cannot be neglected.

For Stephen King's novel, the situation looks quite the same.  On
september 2003, the Board of Directors of the National Book Foundation
announced that its 2003 Medal for Distinguished Contribution to American
Letters would be conferred to Stephen King. This is America's most
prestigious literary prize. The ceremony took place two months later, on
11/19/2003, shortly after the observed peak. As for ``Star Wars,''
the time extension of the exogenous impact of news cannot be
neglected.

\begin{figure}
    \begin{center}
      \includegraphics[width=1\textwidth]{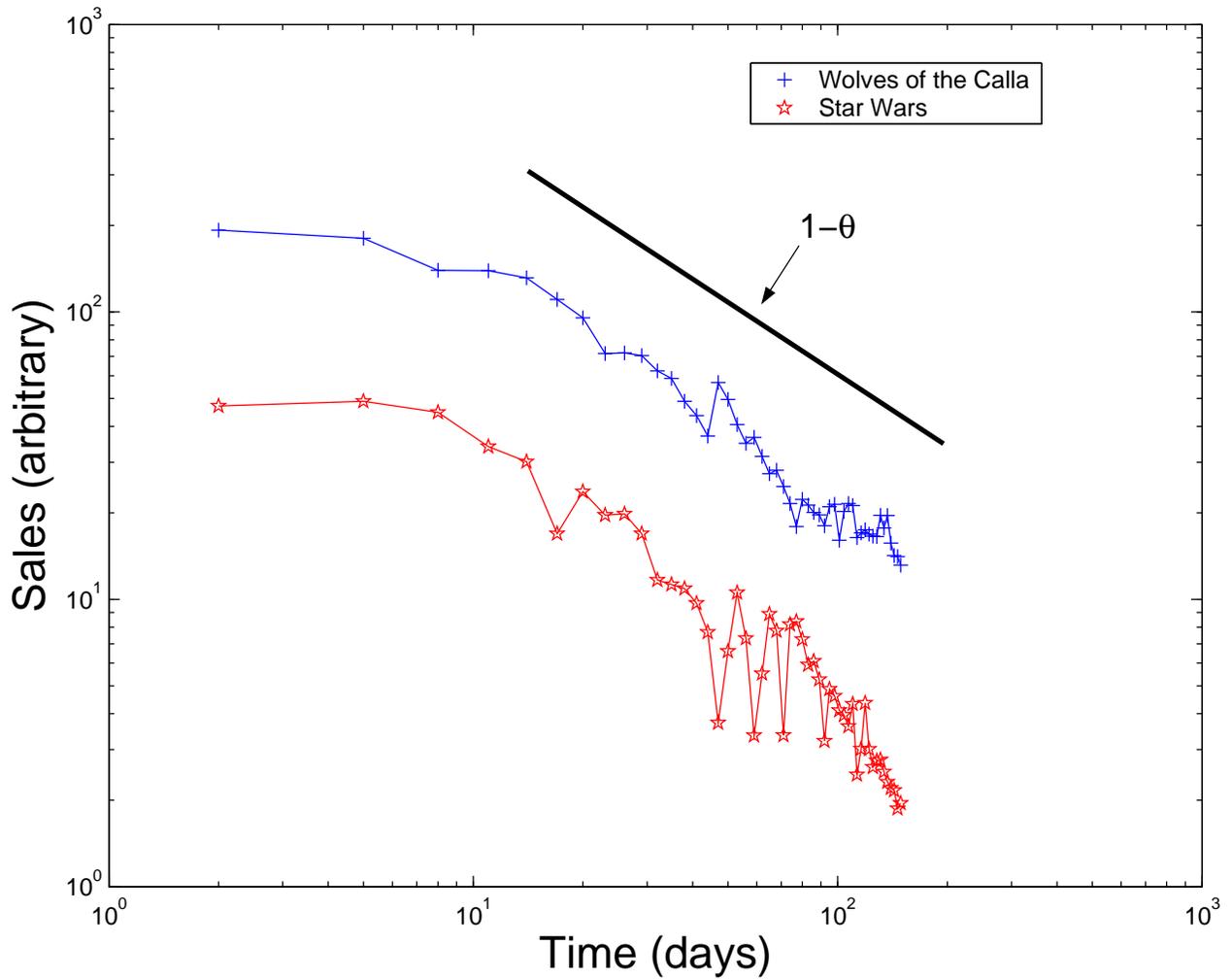}
      \caption{Relaxation of the sales after the
peak for ``Star Wars'' and ``Wolves of the
Calla.''
      The expected power law behavior for exogenous shocks
      can only be observed for $t> 10-20$ days. The small value of the slope for
      $t<10-20$ days can be explained by considering the time duration of
      the external shock triggered by the media as explained in the text.}
      \label{star_wars}
    \end{center}
\end{figure}

This finding suggests that modeling external influence --news,
advertisement-- by a Dirac function $\delta(t)$ is not adequate in some
cases. The external shock may have a significant duration $T$. We then expect
the sales to grow slowly and exhibit a plateau
over a time scale proportional to $T$
and then to crossover to the exogenous
$1/t^{1-\theta}$ decay rate for times $t > T$.
Fig.~\ref{star_wars} shows that this is indeed the case for the
two books ``Star Wars'' and ``Wolves of the Calla'' which were
exceptions to our classification in terms of their acceleration pattern:
the long durations of the peaks are consistent with the fact that the
external news impacted over an extended period of time, leading
to our misclassification as endogenous with respect to their acceleration
before the peak. The time scale of about 20 days of the plateau is
consistent with the known duration of the external news.

These two examples show the need to refine our analysis to allow for a
more general description of external news. In particular, the epidemic
model can in principle be used to invert the amplitude of the flow of
news $\eta(\tau)$ from the time series of the sales $S(t)$. But
to be effective and reliable, this
would require better data, for instance directly working on Amazon sales
rather than on reconstructed sales from ranks.

\subsection{Robustness of the results upon
variation of the conditions of peak selection}

Previously, we kept 30 peaks corresponding to
those which give a fit with a power law
with a correlation coefficient larger than $0.95$
and which have an increase before
the peak of a factor less than $k_{endo}=2$ or of
a factor more than  $k_{exo}=30$.
It is interesting to discuss the results for less restrictive conditions.

\begin{figure}
                \begin{center}
                  \includegraphics[width=0.7\textwidth]{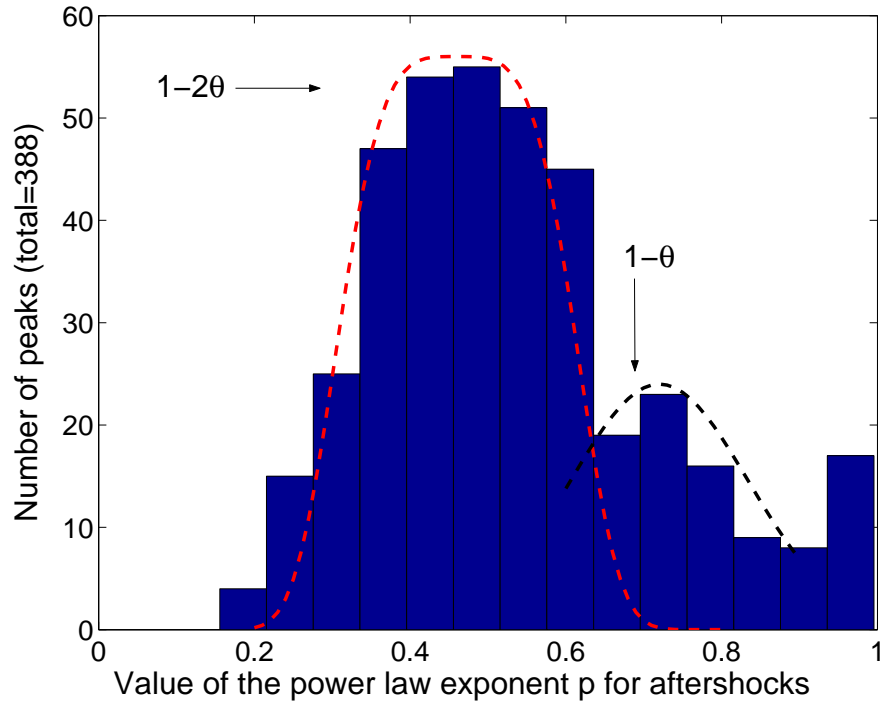}
                  \caption{Histogram of the estimates of the power law
                  exponents $p$ of the relaxations of the sales using
                  a selection criterion different from that of
                  Fig.\ref{histo}: among the 1,013 peaks of our pre-filtered
                  database, we kept those which have a
                  correlation coefficient larger than $0.9$, giving 388 peaks.
                  }
                  \label{aotherpall}
  \end{center}

\end{figure}

Fig.~\ref{aotherpall} is similar to
Fig.~\ref{histo} and shows the distribution of
power law
exponents for the decrease of sales after peaks
for those peaks whose relaxation can
be fitted by a power law with a correlation coefficient larger than $0.9$. This
less restrictive condition on the quality of the
power law fit selects 388 peaks
out of the initial 1013 peaks of our prefiltered data
set, i.e., close to three times
more peaks. We again find two rather clearly defined clusters, which can
be associated with the endogenous and the exogenous classes as previously.

\begin{figure}
                \begin{center}
                  \includegraphics[width=0.7\textwidth]{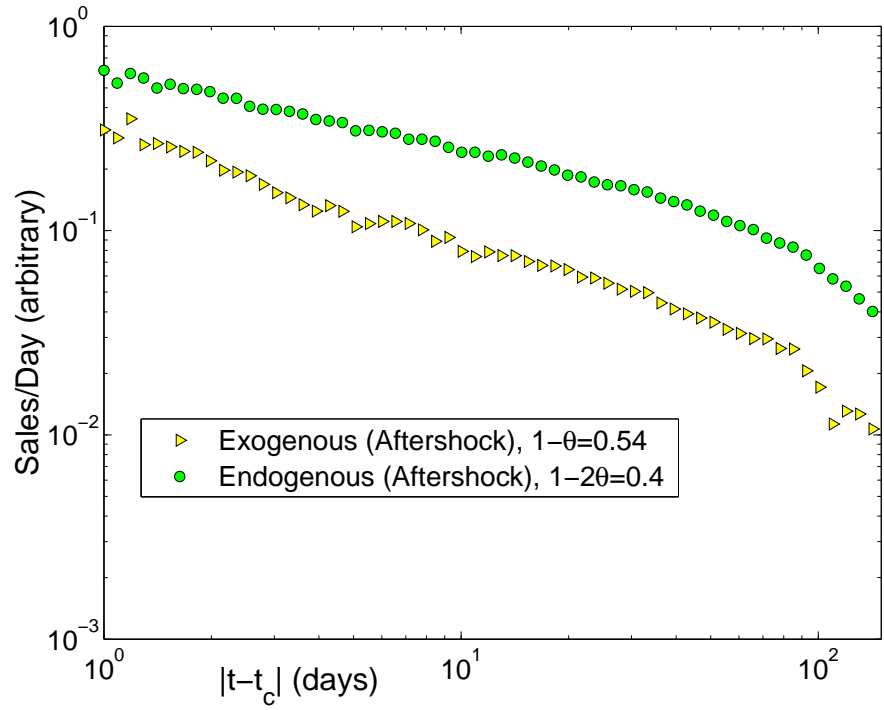}
                  \caption{Relaxation of sales after the peak following the same
                  methodology as for Fig.\ref{stack_after} but with the same set
                  as in Fig.~\ref{aotherpall} with
the clustering procedure explained
                  in the text. }
                  \label{fjkla}
  \end{center}
\end{figure}

Among the 388 peaks that have a correlation coefficient more than $0.9$,
we kept the 270 peaks which have an increase before the peak less than
$k_{endo}=8.5$ or more than $k_{exo}=12.5$. These two coefficients were
selected empirically, such that the algorithm make a distinction between
slow and fast acceleration similar to what common sense would suggest.
The $388-270=118$ rejected peaks correspond to blurred situations where
the acceleration of the sales is neither fast nor progressive. Fig.~\ref{fjkla}
is similar to Fig.~\ref{stack_after} but for these 270 peaks. We again
obtain two different power law slopes, a faster decrease and larger exponent
for the class classified as exogenous with
respect to its fast foreshock acceleration
than for the class classified as endogenous with
respect to its progressive foreshock acceleration.
For ($k_{endo}$,$k_{exo}$)=(8.5,12.5), we obtain
$p=0.54$ for the exogenous class and $p=0.4$
for the endogenous class. To test the sensitivity
with respect to the coefficients
$k_{endo}$ and $k_{exo}$, we report other values.
For ($k_{endo}$,$k_{exo}$)=(5,20) (respectively
(2,30)) which are not shown, we obtain
a mean exponent for the exogenous class equal to
$0.55$ (resp. $0.57$) and $0.39$ (resp. $0.39$)
for
the endogenous class. While the results are qualitatively consistent with those
obtained with more stringent conditions, we see
that the exponent for the exogenous class
is a bit too small. This may be due to the duration of the news which are not
instantaneous as discussed above and to the
existence of other factors and disturbances
not described here.

\section{Further tests and constraints on the
rank-sale conversion (\ref{pdf S})}
\label{Direct measure on rank}

In this section, we compare the prediction of the model with the data
on changes of ranks to test the validity of the
power law distribution (\ref{pdf S})
that we used to convert ranks into sales.

Let us denote $\Delta R(t)
\equiv R(t+\Delta t) - R(t)$ the variation of
rank over the time interval $\Delta t$
(which will be taken fixed and equal to $1$ day).
Let us call $\Delta S(t) \equiv S(t+\Delta t) - S(t)$, the
variation of sales over the same time interval.
It is clear that variations of ranks must be
interpreted relative to past ranks.
A change of a few ranks when a book is selling a
rank 10 is not the same as when
it has rank 10,000. This motivates us to study the conditional rank variation
defined by
\be
\langle \Delta R(R) \rangle \equiv \mathrm{E}[\Delta R (t)|R(t)]~,
\label{mgmls}
\ee
defined as the rank variation from time $t$ to
$t+1$ for a book which was at rank $R(t)$
at time $t$. Similarly, the conditional sale variation
\be
\langle \Delta S(S) \rangle \equiv \mathrm{E}[\Delta S (t)|S(t)]
\ee
is the expectation of the variation of the sales
conditioned on its value before.

First, we will derive $\langle \Delta S (S) \rangle$ from our model.
Then, using the postulated rank-sales conversion given by (\ref{pdf
S}) will provide a prediction for $\langle \Delta R(R) \rangle$. We
will also measure $\langle \Delta R(R) \rangle$ directly without
using the conversion rank-sales, and compare to test the ranks-sales
conversion.

\subsection{Prediction of the model of section \ref{Description of the Model}}
\subsubsection{Derivation of $\langle \Delta S(S) \rangle$}

Let us define
\begin{equation}
\widetilde{S}(t) \equiv \int_{-\infty}^{t} \mathrm{d} \tau
\widetilde{\eta}\,(\tau)\, \kappa(t-\tau),
\end{equation}
with $\widetilde{\eta} \equiv \eta - \langle \eta \rangle$. Then,
$S(t)=\widetilde{S}(t)+ \frac{\langle \eta \rangle}{1-n}$ and
$\Delta S(t)=\Delta \widetilde{S}(t)$.
Let us expand $\Delta S(t)$ :
     \begin{align*}
     \Delta S(t)= &\Delta \widetilde{S}(t) \nonumber\\
     =&\int_{-\infty}^{t} d\tau
\widetilde{\eta}(\tau) (\kappa(t- \tau + \Delta
t)- \kappa(t-\tau)) \nonumber \\
     +&\int_{t}^{t+\Delta t} d\tau \widetilde{\eta}(\tau)
\kappa(t+\Delta t-\tau).
     \end{align*}
Conditioned on the value of $S(t)$, the
expectation of the second term of the r.h.s is
zero because
the conditioning does not affect times posterior to $t$. This implies
  \begin{equation}
    \langle \Delta S(S) \rangle=\int_{-\infty}^{t}
d\tau
\mathrm{E}\left[\widetilde{\eta}(\tau)|S(t)\right]
     (\kappa(t- \tau + \Delta t)- \kappa(t-\tau)) ~.
     \label{delta S}
     \end{equation}
Following the same reasoning as in
section \ref{dynamical properties}, we have
$\forall S(t), ~~\mathrm{E}[\widetilde{\eta}(\tau)|S(t)] \propto
\widetilde{S}(t)\kappa(t-\tau)$. Replacing this expression in
(\ref{delta S}) obtains
     \begin{align}
     \langle \Delta S(S) \rangle &= - \alpha \widetilde{S}\\
                   &=-\alpha (S- \langle S \rangle),
     \label{delta S 2}
     \end{align}
     with
\be
\alpha \simeq - \int_{-\infty}^{t}\mathrm{d}\tau
(\kappa(t-\tau+\Delta t)-\kappa(t-\tau)) ~,
\ee
which is positive because $\kappa$ is a decreasing function.

\subsubsection{Derivation of $\langle \Delta R(R)
\rangle$ from $\langle \Delta S(S) \rangle$
assuming (\ref{pdf S}) to be valid}

Expression (\ref{pdf S}) implies
that the function $R(S)$ is a power law of exponent $-\mu$. Thus,
taking the logarithmic derivative expressed as a finite difference gives
     \begin{equation}
     \frac{\Delta R}{R}= - \mu \frac{\Delta S}{S}~.
     \label{log derivative}
     \end{equation}
     From (\ref{delta S 2}) and (\ref{log derivative}), we get
     \begin{equation}
     \frac{\langle \Delta R(R) \rangle}{R}=\mu
\alpha (1- \frac{ \langle S \rangle}{S})
     \label{delta R delta S}
     \end{equation}
   $S$ can be expressed as a function of $R$ as follows. Starting from
    (\ref{pdf S}), we express the constant $C$ as $C=\mu S_{min}^{\mu}$ from
    the normalization of $p(S)$ in the interval
from a minimum sale $S_{min}$ to infinity.
Then, $\langle S \rangle =\int p(S) ~S~\mathrm{d}S =
\frac{\mu}{1-\mu} S_{min}$. We have assumed $\mu > 1$, which implies
that a majority of the sales are coming from books with large ranks
(i.e., low sales) and not from the few blockbusters in the top
ranks. This leads to \be
    R(S)=N P_{>}(S) =N \biggl(\frac{\mu-1}{\mu}\biggr)^\mu
         \biggl(\frac{\langle S \rangle}{S}\biggr)^\mu~,
    \label{R(S)}
\ee
  where $N$ is the `total' number of books.
It is not very clear which value should be
taken for $N$. Amazon sells and ranks several
millions of books. But, should our
power-law conversion be true, it won't probably be valid for the largest ranks
in the million range. We can probably expect to have $N \approx 10^4$ because
Amazon's change of ranking scheme around $R=10^4$, creating a
natural population of the first 10,000 ranked books.
Or perhaps, it could be $10^5$.

Putting together
(\ref{delta R delta S}) and (\ref{R(S)}) leads to
   \begin{equation}
   \langle \Delta R(R) \rangle = \alpha \mu R \biggl[ 1-\frac{\mu}{\mu-1}
   \biggl(\frac{R}{N}\biggr)^{\frac{1}{\mu}} \biggr]~.
   \label{delta R}
   \end{equation}
Our epidemic model of book buys together with the assumption of a
power law distribution of sales with exponent $\mu$, expression
(\ref{delta R}) predicts that the average variation of ranks is
proportional to the rank itself for small $R$ and to
$-R^{1+\frac{1}{\mu}}$ for large rank. The non-monotonicity of
$\langle \Delta R(R) \rangle$, i.e., $\langle \Delta R(R) \rangle >
0$ for small $R$ and $\langle \Delta R(R) \rangle<0$ for large $R$
simply reflects that best-sellers tend to lose ranks because being a
best-seller requires to have continuously sources $\eta(t)$'s of buyers
above the average, which can only be achieved for a relatively short
time. Conversely, poorly ranked books can only improve their ranking
on average.

\subsection{Empirical test}

\begin{figure}
                \begin{center}
                  \includegraphics[width=1\textwidth]{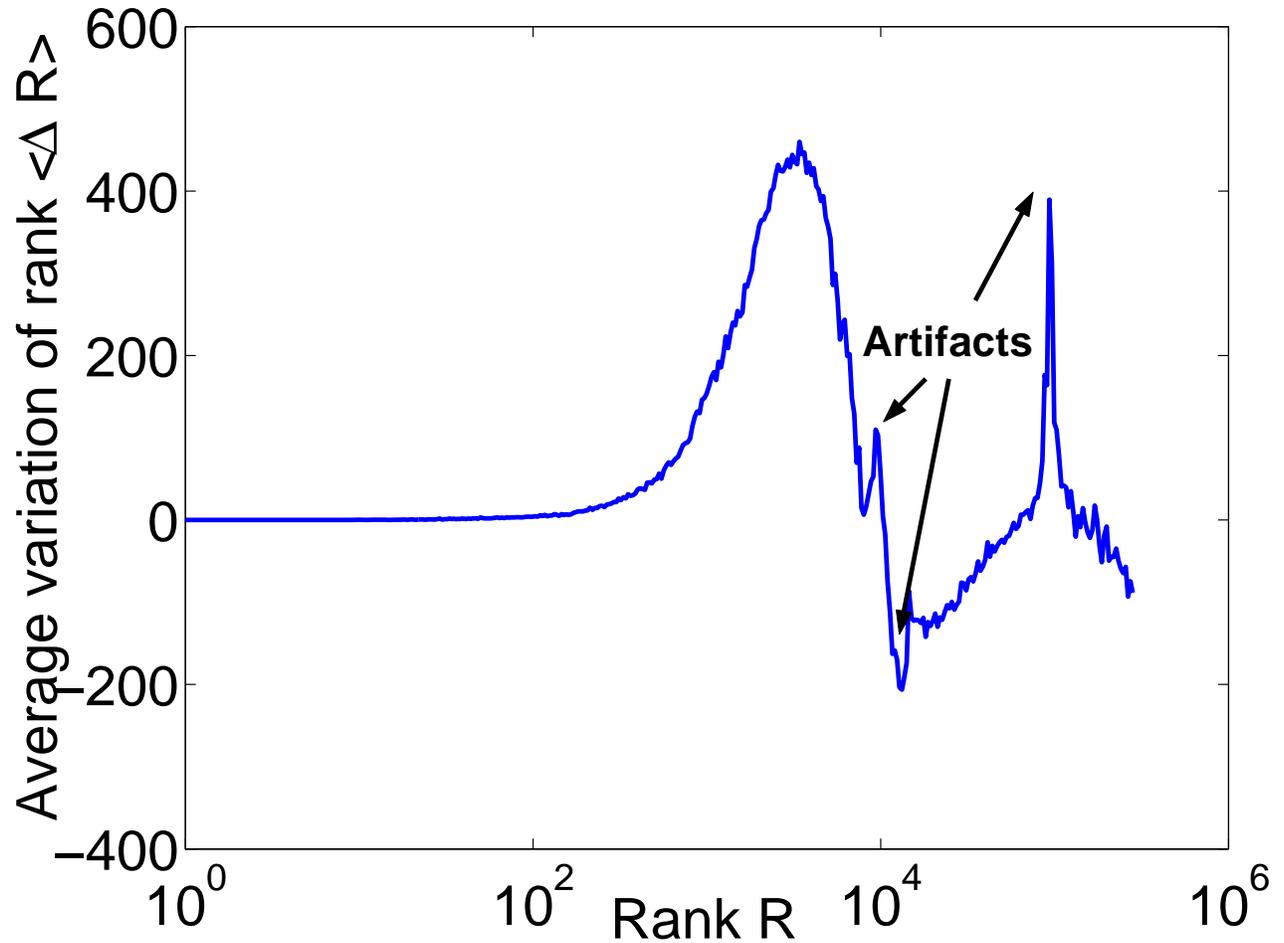}
                  \caption{Average variation
$\langle \Delta R \rangle$ of the rank in
function of
                  the rank itself.
                  This measure was performed over a sample of 14000 books.
One can note spurious behaviors for ranks close to $10^4$ and
$10^5$. This can be rationalized by a shift in ranking schemes for
these value (see section \ref{official statement}). The data can
only be exploited for $R<10^4$ to avoid these artifacts. The most
striking feature is the non-monotonous behavior of $\langle \Delta R
\rangle$ for $R<10^4$. This enables us to reject an exponential
conversion (see section \ref{toward}).}
                  \label{overall_delta_R}
                  \end{center}
\end{figure}

Fig.\ref{overall_delta_R} shows the overall
behavior of $\langle \Delta R(R) \rangle$. For our purpose, we will
ignore ranks $R > 10^4$ because, for such ranks, the Amazon
ranking scheme is not appropriate as already discussed in section
\ref{ranking schemes} and illustrated by the artifacts
for $R \approx 10^4$ and $R \approx 10^5$. These spurious peaks
reflect the shift of Amazon between different ranking schemes.

\begin{figure}
                \begin{center}
                  \includegraphics[width=1\textwidth]{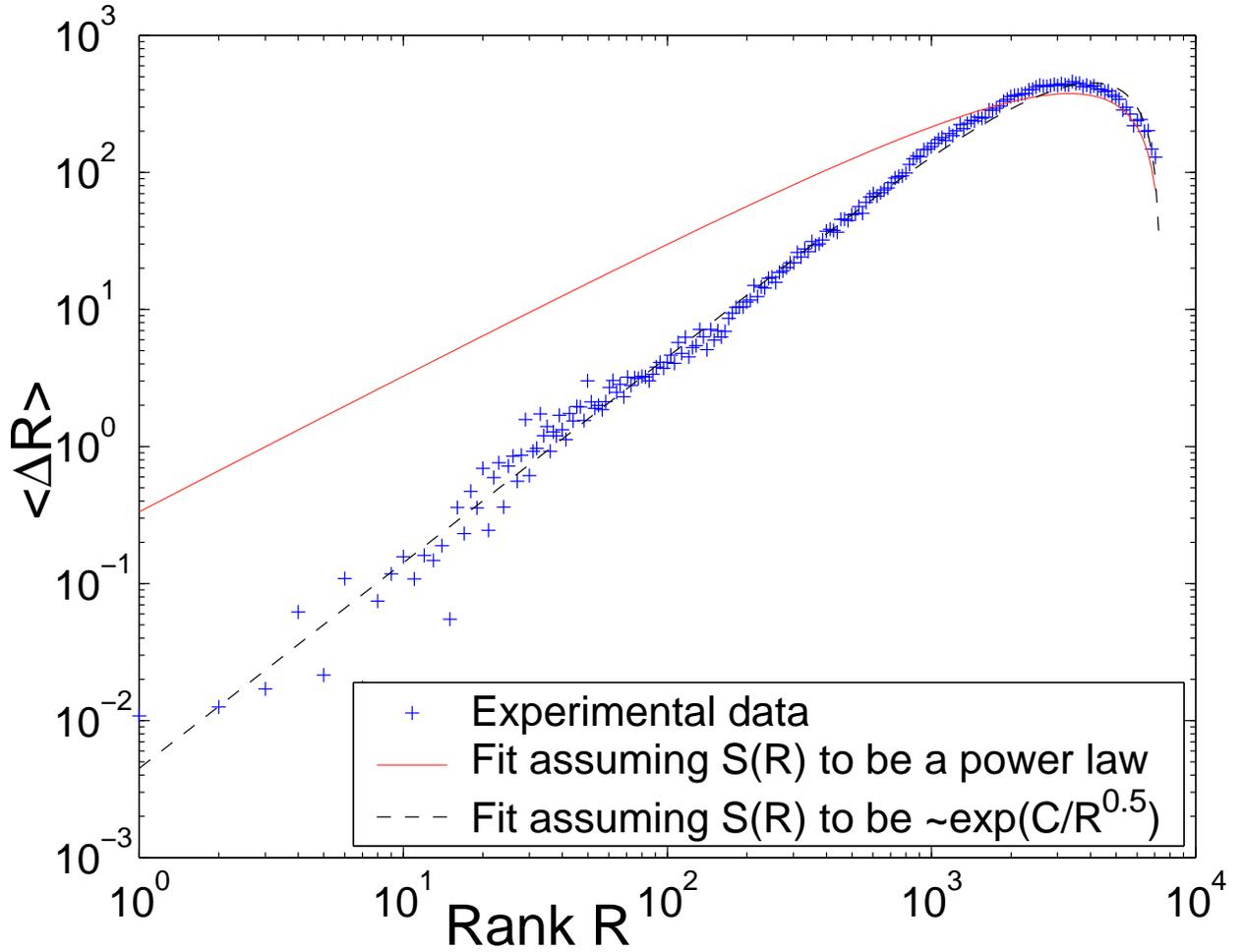}
                  \caption{Average variation of
the rank $\langle \Delta R \rangle$ as a function
of
                  the rank itself. We only take
into account those time series that do not exhibit
                    abrupt peaks (i.e., we discard exogenous shocks).}
                    \textbf{plus} : experimental data,
  \textbf{solid line} : fit assuming $S(R)$ to be a power law,
\textbf{dashed line} : fit assuming $S(R)$ to be
$S(R)\propto \exp(\frac{C}{\sqrt{R}})$
                  \label{loglog_Delta_R}
                  \end{center}
\end{figure}

The part of Fig.\ref{overall_delta_R} for $R < 10^4$ is magnified in
Fig.\ref{loglog_Delta_R}. The predicted non-monotonous behavior is
observed but expression (\ref{delta R}) do not fit the experimental
data for the small ranks. The log-log plot of
Fig.\ref{loglog_Delta_R} clearly shows that $\langle \Delta R
\rangle$ is given by \be \langle \Delta R \rangle  \sim
R^\beta~,~~~{\rm with} ~\beta=1.5 \pm 0.05~, \label{ngutjs} \ee over
approximately three decades. Thus, $\langle \Delta R \rangle$ is not
proportional to $R$ since $\beta$ is significantly larger than $1$.
This departure from linearity can either signal a problem
with the prediction (\ref{delta S 2}) of the model or can be
due to a breakdown of the power law assumption (\ref{pdf S}).

\subsubsection{First option: breakdown of
$\mathrm{E}[\widetilde{\eta}(\tau)|S(t)] \propto
\widetilde{S}(t)$}

Suppose instead of (\ref{delta S 2}) that \be \langle \Delta S(S)
\rangle \sim - \widetilde{S}^x~, \label{gnvjelel} \ee where $x$ may
be different from $1$. Then, following step by step the derivation
leading to (\ref{delta R}), we obtain \be \langle \Delta R(R)
\rangle \sim R^{1-x \over \mu}~R~, \label{mgmldls} \ee where we have
used the power law conversion (\ref{pdfranksales}) between sales and
ranks derived from (\ref{pdf S}). When (\ref{delta S 2}) holds,
$x=1$ and expression (\ref{mgmldls}) recovers the linear dependence
of $\langle \Delta R(R) \rangle$ with $R$, for not too large $R$,
quantified by the first term in the r.h.s. of (\ref{delta R}). Now,
expression (\ref{mgmldls}) is compatible with the data
(\ref{ngutjs}) together with the measurement $\mu \simeq 2$ only if
we take $x=0$, which implies that $\langle \Delta S(S) \rangle$ does
not depend in a first approximation on $\widetilde{S}$. This in turn
implies that the result $\mathrm{E}[\widetilde{\eta}(\tau)|S(t)]
\propto \widetilde{S}(t)\kappa(t-\tau)$ obtained above following the
reasoning of section \ref{dynamical properties} does not hold and
must be replaced by the statement that
$\mathrm{E}[\widetilde{\eta}(\tau)|S(t)]$ is independent of
$\widetilde{S}(t)$. Note that the dependence of
$\mathrm{E}[\widetilde{\eta}(\tau)|S(t)]$ on $\kappa(t-\tau)$ is not
necessarily linked to the validity of
$\mathrm{E}[\widetilde{\eta}(\tau)|S(t)] \propto \widetilde{S}(t)$
and thus the proportionality
$\mathrm{E}[\widetilde{\eta}(\tau)|S(t)] \propto \kappa(t-\tau)$ can
still hold, ensuring the validity of the exponent $1-2\theta$ for
the foreshock and aftershock sales associated with an endogenous
peaks (see Table \ref{distinguishing shock}).

A possible interpretation of the breakdown of
$\mathrm{E}[\widetilde{\eta}(\tau)|S(t)] \propto \widetilde{S}(t)$
is that the amplitude of the sales have a large stochasticity from book to book
and the dependence of precursory innovations on
the future amplitude of the sales's peak
is lost. In sum, if we accept the conclusion
drawn in insight that the sales' innovations
are weakly dependent or are actually independent
of the amplitude of the sales' peak, then
the observed law (\ref{ngutjs}) can be seen as a dramatic confirmation of the
power law  (\ref{pdf S}) through the rank-to-sale
power law conversion (\ref{pdfranksales})
in the range of ranks up to a few thousands. However, as a note of caution,
we cannot exclude the possibility that $x \neq 0$ and $\mu \neq 2$
as long as $(1-x)/\mu=1/2$ holds: in such a case, expression (\ref{mgmldls})
shows that we would recover the observed power law relationship (\ref{ngutjs})
from the analogous derivation that led to (\ref{delta R}).

\subsubsection{Second option: deviations from the power law conversion}
\label{toward}

Let us consider a general function $S(R)$ relating ranks to sales.
Equation (\ref{log derivative}) becomes $\Delta S=S^\prime(R)\Delta
R $. Using the empirical evidence $\langle \Delta R \rangle \propto
R^\beta$ and assuming still the validity of $\langle \Delta S
\rangle \propto - S$ for small $R$, we get
$S(R)=S_{min}\exp(\frac{C}{R^{\beta-1}})$, where $S_{min}$ and $C$
are two constants. Assuming that this expression for small $R$ can
be used for all ranks of interest, we get :
\begin{equation}
\langle \Delta R(R) \rangle= \frac{\alpha R^\beta}{C (\beta-1)} \biggl[
1-\frac{<\!\!S\!\!>}{S_{min}}
\exp\biggl(-\frac{C}{R^{\beta-1}}\biggr)\biggr].
\label{delta R exp(sqrt)}
\end{equation}
The two constants  $S_{min}$ and $C$ can be
adjusted to provide a rather good fit by
(\ref{delta R exp(sqrt)}) to the data, as shown in Figure \ref{loglog_Delta_R}.
By construction, $\langle \Delta R \rangle$ has the
correct behavior for small $R$. All the difficulty is to describe how
$\langle \Delta R \rangle$ reaches its maximum and then decreases.
Typically, the fit gives $C=165$, which leads to
$\frac{S(1)}{S_{min}}=\exp(165)\approx 10^{70}$. Obviously, this
is totally absurd ! Our mistake was to derive $S(R)$ for small $R$
in order to use it to calculate the behavior of
$\langle \Delta R \rangle$ for any $R$.

A better approach is to invert the empirical function $\langle
\Delta R(R) \rangle$ to get $S(R)$, without making assumptions about
it. For a given function $S(R)$, equation (\ref{delta R delta S})
becomes
\begin{equation}
S^\prime(R) \langle \Delta R(R) \rangle=-\alpha (S(R)- \langle S \rangle~.
\label{delta R delta S general}
\end{equation}
Thus, we can express $S(R)$
as a function of $\langle \Delta R \rangle$ and obtain
\begin{equation}
S(R) = \langle S \rangle +
(S_{\mathrm{max}}- \langle S
\rangle)\exp\biggl(-\alpha\int_{1}^{R}\frac{1}{\langle
\Delta
R(R^\prime) \rangle} \mathrm{d}R^\prime\biggr)~,
\label{final_equ}
\end{equation}
where $S_{max}$ is a constant equal to $S(R=1)$.

\begin{figure}
                \begin{center}
                  \includegraphics[width=\textwidth]{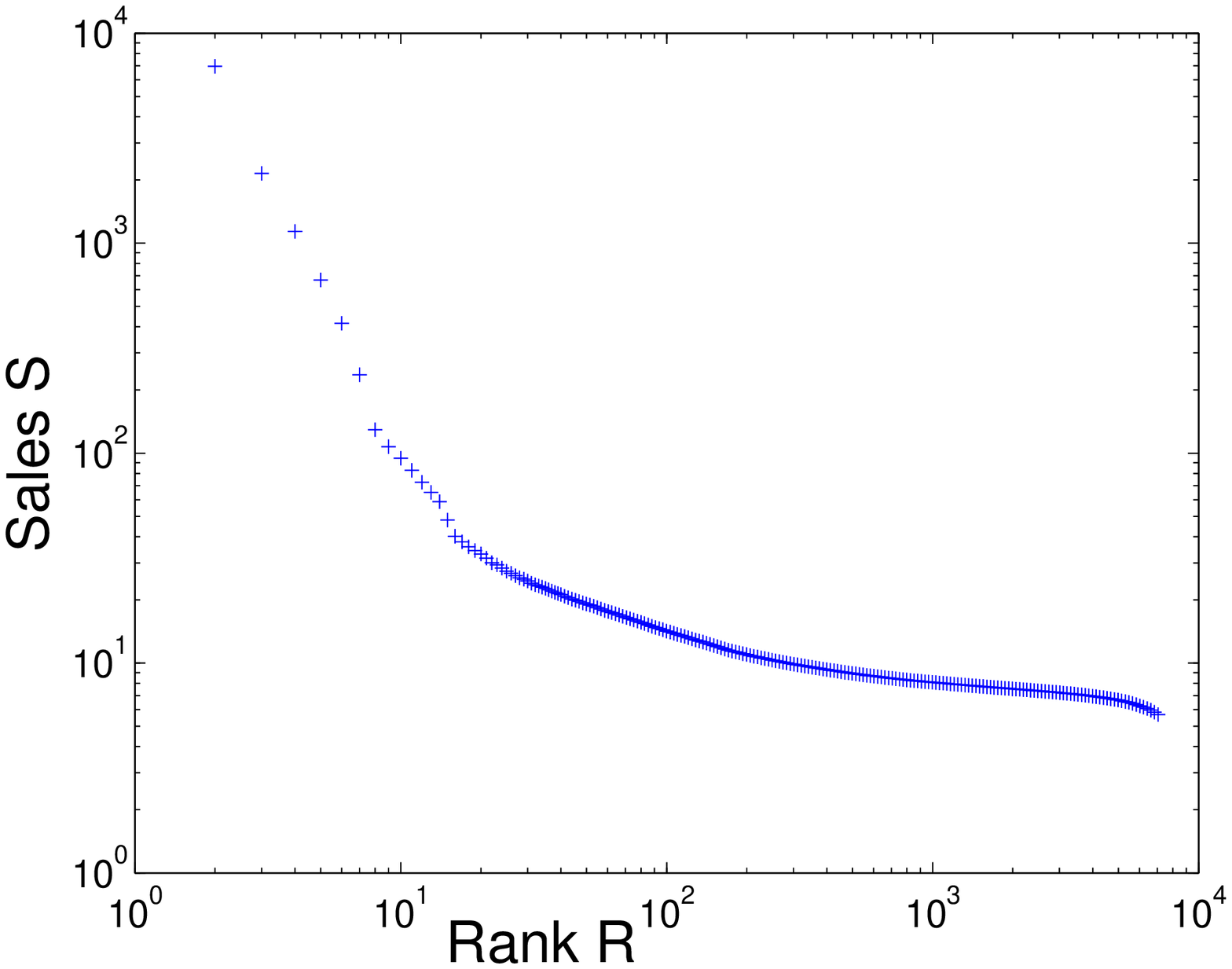}
                  \caption{Rank-to-Sales
conversion function $S(R)$ estimated using
                  expression (\ref{final_equ}) for arbitrary
                  values of $\alpha$, $\langle S
\rangle$, $S_{max}$. Consequently, the units on
the y-axis
                  are arbitrary.}
                  \label{conversion_exp}
                  \end{center}
\end{figure}

The problem to use (\ref{final_equ}) is that we do not know the value
of the constant $\alpha$, which is rather crucial as it
appears in the exponential. Nevertheless, whatever the value of
$\alpha$, we observe roughly the same behavior for $S(R)$.
Fig.\ref{conversion_exp} shows $S(R)$ obtained by inserting the
empirical dependence $\langle \Delta R(R) \rangle$ as a function of $R$ in
expression (\ref{final_equ}) for fixed values of
$\alpha$, $S_{max}$ and $\langle S \rangle$. The
``fast'' decrease for small $R$
(typically $R<10-20$) followed by a ``slow'' one for large $R$'s
is typical of most reasonable parameters. This shows that the conversion can be
roughly seen as a power law (an approximate straight line in a log-log plot)
in the range $R \in [20,10^4]$. But we are
not able to determine the value of the exponent
from this approach because we don't know
$\alpha$.

In summary, the direct determination of the function $S(R)$ requires
some additional assumption as shown in the various attempts
developed above. Nevertheless, we have seen that a power law would
explain the observed non-monotonous behavior for $\langle \Delta
R(R) \rangle$. Notice that if we assume $S(R)$ to be exponential :
$S(R)=S_0\exp(-R/R_\star)$, the relation (\ref{delta R delta S
general}) implies :
\begin{equation}
\langle \Delta R(R) \rangle= \alpha
R_\star\biggl(1- \frac{\langle S \rangle}{S_0}
\exp\bigl(\frac{-R}{R_\star}\bigr) \biggr),
\end{equation}
which is obviously monotonous and can thus be
rejected by comparison with the data.

\section{Conclusion}

In our study of the ranks of books sold by Amazon.com, we
    have shown that sales shocks can be classified into two
    categories: endogenous and exogenous.
    We have used two independent ways of classifying peaks, one based on the
    acceleration pattern of sales (see section \ref{Distribution of exponents})
    and the other based on the exponent of the relaxation (see
    section \ref{Stacking the peaks}). We have developed an epidemic model
    of influences between buyers connected within a network of acquaintances.
    The comparison between the predictions of the model
    and the empirical data suggests that social
networks have evolved to converge
   very close to criticality (here in the sense of critical branching
   processes). It means that tiny perturbations, which in
   any other state would be felt only locally, can propagate almost
   without any bound. Studies of critical phenomena shows that very
   different systems can exhibit fundamental similarities, this is
   generally referred as universality.

While we have
   emphasized the distinction between exogenous and endogenous
    peaks
    to set the fundamentals for a general study, we also find closely
    repeating peaks as well as peaks that may not be pure members of a
    single class. In a sense, there are no real ``endogenous peaks'', one
    could argue, because there is always a source or a string of news
    impacting on the network of buyers. We have thus distinguished
    between two extremes, the very large news impact and the
    structureless flow of small news amplified by the cascade effect
    within the network. One can imagine and actually observe a continuum
    between these two extremes, with feedbacks between the development of
    endogenous peaks and the attraction of interest of the media as a
    consequence, feeding back and providing a kind of exogenous boost,
    and so on. Our framework allows us to generalize beyond these two
    classes and to predict the sales dynamics as a function of an
    arbitrary set of external sources.

If Amazon.com would release its data, we suggest future promising directions
of inquiries.
    \begin{itemize}
     \item Before anything else, the same study   should be
     done again. We expect more accurate results as we work with the real
     sales and not only with an estimate derived from the ranks.
     \item Secondly, a direct access to sales should make it possible to
     reverse $S(t)$ to get access to the news
innovations $\eta(t)$ describing the sources of
     spontaneous buys. We currently describe $\eta(t)$ as a white noise
     distributed according to a power law. This zero order approximation
     could be improved and leads to a better understanding of the statistical
     properties of the news $\eta(t)$. One can even imagine to reconstruct
     the specific history of news that were amplified by the epidemic process
     to obtain a given sales history for a given book.
     \item Different kinds of books should involve different
     social networks. Take the example of the book ``Divine
     Secrets of the Ya-Ya Sisterhood'' by R. Wells. It became a bestseller
     two years after publication, with no major advertising campaign
     \cite{tipping point}. Following the reading of this originally small
     budget book, ``Women began forming \textsl{Ya-Ya} Sisterhood groups
     of their own [...]. The word about \textsl{Ya-Ya} was spreading
     [...] from reading group to reading group, from \textsl{Ya-Ya} group
     to \textsl{Ya-Ya} group''\cite{tipping point}. By looking at
     different classes of books, we would expect to highlight different
     network characteristics \cite{6 degrees,Linked}.
    \item Amazon has the addresses of its customers. This can be used to
    study the geographical spread of books.
    \end{itemize}

\begin{figure}
                \begin{center}
                  \includegraphics[width=0.7\textwidth]{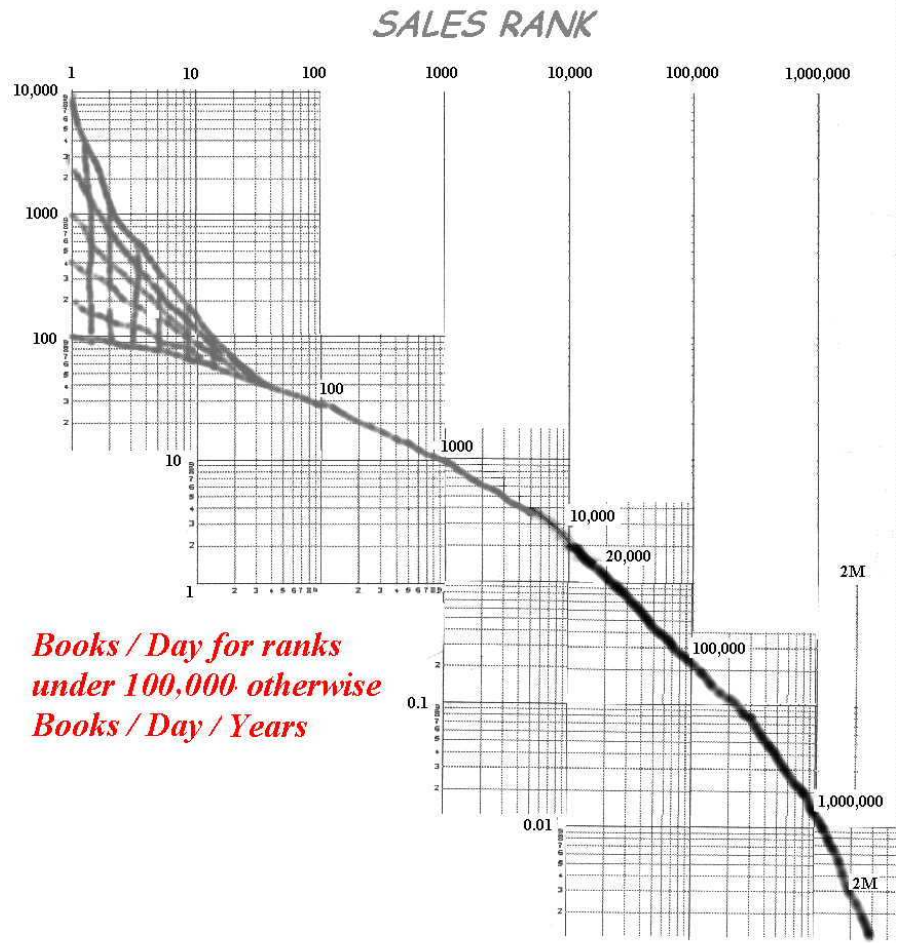}
                  \caption{Rank ordering (Zipf
plot) as in Fig.~\protect\ref{rosenthal}
                  presented in
\protect\cite{surfing amazon}, extended to the
smallest as well as largest ranks. See these two extensions.
Reproduced with kind permission from M. Rosenthal.}
        \label{rosenthal2}
                \end{center}
    \end{figure}

{\bf Acknowledgements}: The present analysis of Amazon sales ranks is
unauthorized and in no way sponsored by Amazon.
We thank Y. Ageon and T. Gilbert 
for help and useful discussions and M. Rosenthal for important 
information on the Amazon ranking system. We also thank M. Rosenthal
for his permission to reproduce figure \ref{rosenthal2}.

\clearpage

\clearpage

\clearpage

\clearpage

\clearpage

\clearpage

\clearpage

\clearpage

\clearpage

\clearpage

\clearpage

\clearpage

\clearpage

\clearpage

\clearpage

\clearpage

\begin{thebibliography}{99}

\bibitem{ruelle} Ruelle, D.,
Physics Today, 57(5), 48-53 (2004).

\bibitem{reviewsorendoexo} Sornette, D.,
Endogenous versus exogenous origins of crises,
in monograph on extreme events, V. Jentsch editor (Springer, 2005)
(http://arxiv.org/abs/physics/0412026)

\bibitem{endoPRL} Sornette, D., F. Desch\^atres, T. Gilbert and Y. Ageon,
Phys. Rev. Letts. 93 (22), 228701 (2004).

\bibitem{Roehneretal} Roehner, B.M., D. Sornette and J.V. Andersen,
Int. J. Mod. Phys. C 15 (6), 809-834 (2004).

\bibitem{Sorfirstbook} D. Sornette,
\textsl{Critical Phenomena in Natural Sciences},
Springer Series in Synergetics, 2nd ed (2004)

\bibitem{surfing amazon} Rosenthal, M., http://www.fonerbooks.com/surfing.htm

\bibitem{system with memory} D. Sornette and  A. Helmstetter,
Physica A, 318, 577-591 (2003).

\bibitem{RoehnerSorfrenzy} Roehner, B. M. and D. Sornette,
European Physical Journal B 16, 729-739 (2000).

\bibitem{new york times} J. Brody, Push up the
weights, and roll back the years,
\textsl{The New York Times} \textbf{F} 7 (June 4, 2002).

\bibitem{6 degrees} D.J. Watts, \textsl{Six
Degree}, Published by W.W. Norton \& Company
(February 2003)

\bibitem{Linked} A.L. Barab\`{a}si, \textsl{Linked} (Perseus, Cambridge,
2002).

\bibitem{tipping point} M.Gladwell, \textsl{The tipping point : how little
things can make a big difference} (Back Bay Books, 2002).

\end{thebibliography}
\end{document}